\documentclass[12pt]{iopart}

\usepackage{iopams}  
\usepackage{color}
\usepackage{subcaption}
\usepackage{siunitx}
\usepackage{graphicx}
\usepackage{booktabs}
\begin{document}

\title[]{Roles of non-axisymmetric perturbations in free drift vertical displacement events on EAST}

\author{Haolong Li$^{1}$, Ping Zhu$^{2,3*}$, Hang Li$^{4*}$, Muquan Wu$^{4}$, Xiang Zhu$^{4}$, and Jingting Luo$^{1}$}

\address{1. College of Physics and Optoelectronic Engineering, Shenzhen University, Shenzhen 518060, China}
\address{2. State Key Laboratory of Advanced Electromagnetic Technology, International Joint Research Laboratory of Magnetic Confinement Fusion and Plasma Physics, School of Electrical and Electronic Engineering, Huazhong University of Science and Technology, Wuhan, 430074, China}

\address{3. Department of Nuclear Engineering and Engineering Physics, University of Wisconsin-Madison, Madison, Wisconsin 53706, United States of America}
\address{4. Advanced Energy Research Center, Shenzhen University, Shenzhen 518060, China}
\ead{zhup@hust.edu.cn, lih@szu.edu.cn}
\vspace{10pt}

\begin{abstract}

The safe operation of most tokamaks, especially the largen sized ones, rely on the feedback control of the vertical displacement events (VDEs). However, most these feedback control systems are based on the axisymmetric VDE models. In this work, we use NIMROD simulations to study the roles of non-axisymmetric perturbations in free drift vertical displacement events on EAST. The high-$n$ modes in non-axisymmetric VDE grow first, which drive the formation of high-$n$ magnetic island chains. Subsequently, the magnetic island chains grow and overlap with each other, leading to the destruction of the magnetic flux surface, which induces a minor disruption and accelerates the start of the following major disruption. The magnetic island and the stochastic magnetic field allow the toroidally asymmetric poloidal plasma current to jet towards the hoop force direction, forming the finger and filamentary structures. Such a plasma current asymmetry strongly depends on the anisotropy in thermal transport coefficients.

\end{abstract}

%
\vspace{2pc}
\noindent{\it Keywords}: tokamak disruption, MHD simulation, vertical displacement events, rotation of asymmetry
%

\submitto{\NF}
%
\maketitle
%
%

\section{Introduction}

For higher confinement and the benefits to the impurity elimination~\cite{Senichenkov_2019}, the elongated cross-sections have been adopted by most advanced tokamaks and the next generation reactors (e.g. ITER~\cite{hender_chapter_2007} and CFETR~\cite{wan_overview_2017}). An elongated tokamak plasma is unstable to toroidally axisymmetric mode~\cite{Mukhovatov_1971}, which if left uncontrolled, would allow the plasma to move upwards or downwards, and eventually strike the vacuum vessel, triggering a rapid quench of the plasma pressure and current, a scenario known as the Vertical Displacement Event (VDE). In addition, asymmetries induced by VDEs can lead to horizontal forces on electrically conducting structures~\cite{WALKER2009665,Gruber_1993,Strait_1991}, and the forcing may rotate at rates that are comparable to mechanical harmonics~\cite{Neyatani_1999}. Therefore, comprehensive analysis, including large-scale simulations seeking to identify the worst VDE cases, is necessary as part of the design process.

Early theory by Mukhovatov and Shafranov approximates the tokamak plasma as a current-carrying loop to derive the linear stability criterion for a large aspect-ratio tori~\cite{Mukhovatov_1971}. Later, more realistic toroidal models of the plasma region including elliptically shaped cross-sections with parabolic~\cite{Haas_1975} and step-function~\cite{Fitzpatrick2008A} pressure profiles are applied into the linear ideal-plasma analysis for vertical instability. The separation of time scales between resisitive dissipation, plasma motion, and the Alfv{\'e}nic propagation implies that magnetic topology and force-balance are preserved~\cite{2012Theory,Allen2015Characteristic}, which is applied to investigating marginally stable vertical displacement, including the open-field halo current~\cite{Strait_1991} by Fitzpatrick~\cite{2009Fit}. Including 3D external kink modes, the tokamak-MHD (TMHD) model with a non-ideal wall has been developed for these conditions~\cite{2015PhPl}, and used to investigate the wall-touching kink mode (WTKM)~\cite{2008PhPl}. Meanwhile, Artola has proposed a simple analytical approach to the explanation of the edge safety factor drops based on the conservation of poloidal flux~\cite{artola_understanding_2020}.

In experiments, the Experimental Advanced Superconducting Tokamak (EAST) campaign has observed Hiro currents with direction opposite to the plasma current~\cite{xiong_first_2015,gao_key_2017}, which is consistent with the sign prediction of the WTKM theory~\cite{2008PhPl,2012PhPl}. Later, with ‘ITER-like’ tungsten divertor and W-Like’ graphite divertor, it is observed in EAST that the halo current decreases with the smaller vertical displacement~\cite{chen_characterization_2020}. A new vertical control sheme based on linearized model  and equilibrium modification has been applied to the TCV tokamak, which is is adaptable to different plasma equilibria~\cite{tcv_exp}. And the MGI experiment to intentionally trigger VDE and disruption has been conducted on DIII-D, which indicates the need for sufficiently early MGI triggering in order to achieve effective VDE heat load mitigation~\cite{2015PhPlsss}. The VDE experiments on JET show that the toroidal asymmetry of plasma current induced by VDE is of serious concern~\cite{gerasimov_plasma_2014,Gerasimov_2020}. A scaling law has been obtained for the rotation frequency of the asymmetric Halo current component as a function of toroidal field strength and plasma minor radius ($f_{rot} \sim 1/B_{T} a^{2}$)~\cite{saperstein_halo_2022}.


Most previous simulations on VDE are axisymmetric, such as those using TSC (Tokamak Simulation Code)~\cite{1986JCoPh,Kaye_1994,Nakamura_1996,8357948,Guo_2012}. The CarMa0NL code solves the axisymmetric nonlinear evolutionary equilibrium equations, and when self-consistently coupled to eddy currents equations, is able to describe the 3D wall structure~\cite{Villone_2013, Chen_2019}. Meanwhile, DINA models the tokamak plasma evolution with a series of axisymmetric equilibria and one-dimensional transport~\cite{1993JCoPh}, which has been applied to many fusion experimental devices such as MAST~\cite{2003The}, TCV~\cite{tcv1,tcv2}, and DIII-D~\cite{inproceedings-diiid}.

The assumption of axisymmetry precludes the three dimensional physics, including the 3D magnetic reconnection and the toroidal peak of plasma current density, as well as the three dimensional consequences of VDE process, such as halo current, wall force, and runaway electron, etc. In recent years, several 3D MHD codes have been developed and applied to the simulations on asymmetric VDEs. Based on the realistic equilibrium, M3D~\cite{1999PhPl} simulation results on VDE including TPF and halo fraction are found consistent with experiments~\cite{paccagnella_vertical_2005}. The M3D simulation by Strauss implies that the asymmetric wall force depends on the ratio of the current quench time to the resistive wall penetration time~\cite{2018PhPl}. Pfefferl{\'e} \emph{et al} have applied M3D-C1~\cite{2016PhPl} to the study on the VDE halo/eddy current and the physics processes when plasma touches the wall~\cite{pfefferle_modelling_2018}. JOREK~\cite{Huysmans_2007,Hlzl_2012} has been used to study the VDE halo region on COMPAS~\cite{Artola_2021,artolasuch}. Moreover, Sovinec and Bunkers has developed and applied NIMROD~\cite{nimrod} to the simulation on the forced VDE and the influence from boundary condition~\cite{sovinec2019,bunkers_influence_2020}. Except for highly idealized cases, there are no analytical solutions for events that involve chaotic magnetic field lines and large vertical displacements~\cite{benchmark_3D}. Instead, Krebs \emph{et al} and Artola \emph{et al} have conducted the benchmark of M3D-C1, NIMROD, and JOREK for axisymmetric~\cite{benchmark_2D} and asymmetric~\cite{benchmark_3D} simulations of VDEs.

In this work, based on an EAST-like equilibrium, non-axisymmetric VDE nonlinear simulations including a resistive wall have been performed using NIMROD. The linear stability of the initial equilibrium is investigated. The behaviour of asymmetric toroidal modes are analysed. And the mechanism how the asymmetric toroidal modes destroy the closed flux surface is studied. The rotation of the asymmetry of plasma current is observed, and its origin has been discussed. The goal of this work is to analyse the effects of asymmetry on the free drift VDE process.

The remainder of this paper is organized as follows. Section \ref{set-up} briefly describes the model and computional methods of NIMROD code, the initial reconstructed equilibrium, and some of the key physical parameters used in the simulations. Section \ref{results} presents the linear and nonlinear simulation results on the VDE process, including the destruction and healing of magnetic flux surface, and the asymmetry in plasma current density distribution. Section \ref{summary} provides a summary and discussion on this work and the next steps.

\section{Simulation model and setup}
\label{set-up}


In this work, we use the extended full-MHD equations implemented in NIMROD code~\cite{nimrod}

\begin{eqnarray}
\label{eqn:continue}
		&\frac{\partial n}{\partial t} + \nabla \cdot (n \vec{V}) = \nabla \cdot \left( D_{n} \nabla n  \right) \\
		\label{eqn:mom_eq}
		&mn \left( \frac{\partial}{\partial t} + \vec{V} \cdot \nabla \right) \vec{V} = \vec{J} \times \vec{B} - \nabla (2nT) - \nabla \cdot \underline{\Pi} \\
		\label{eqn:temp_eqn}
		&\frac{n}{\Gamma - 1} \left( \frac{\partial}{\partial t} T + \vec{V} \cdot \nabla T \right) = - nT \nabla \cdot \vec{V} - \nabla \cdot \vec{q}
\\
\label{eqn:Be_eval}
		&\frac{\partial \vec{B}}{\partial t} = - \nabla \times \left( \eta \vec{J} - \vec{V} \times \vec{B} \right)
\\
\label{eqn:Ja_eval}
		&\mu_{0} \vec{J} = \nabla \times \vec{B} \\
		&\vec{q}= -n \left[ \left( \chi_{\parallel} -\chi_{\perp} \right) \hat{b} \hat{b} \cdot \nabla T + \chi_{\perp} \nabla T \right]
\label{eqn:q_transf}
\end{eqnarray}
where $n$ denotes the number density of plasma ($n=n_{i}=n_{e}$), $\vec{V}$ the center-of-mass plasma flow velocity, $m$ the mass of particle, $\vec{J}$ the current density, $\vec{B}$ the magnetic field, $\mu_{0}$ the permeability of vacuum, $T$ the plasma temperature ($T=T_{i}=T_{e}$), $P=2nT$ the total pressure, and adiabatic index $\Gamma=5/3$. The Spitzer model for plasma resistivity $\eta=\eta_{0}(T_{0}/T)^{3/2}$ is adopted, where the $T_{0}$ stands for the plasma temperature at magnetic axis. The viscous stress included in this work is $ \underline{\Pi}=\rho \nu_{\rm{kin}} \nabla \vec{V}$ where $\nu_{\rm{kin}}$ denotes the kinematic viscous coefficient. The anisotropic thermal conduction is modeled in Eq.(\ref{eqn:q_transf}), where $\chi_{\parallel}$ ($\chi_{\perp}$) denotes the thermal transport coefficient parallel (perpendicular) to magnetic field, and $\hat{b}$ the unit vector along the magnetic field.

\begin{figure}[htb]
  \centering
  \includegraphics[width=1\textwidth]{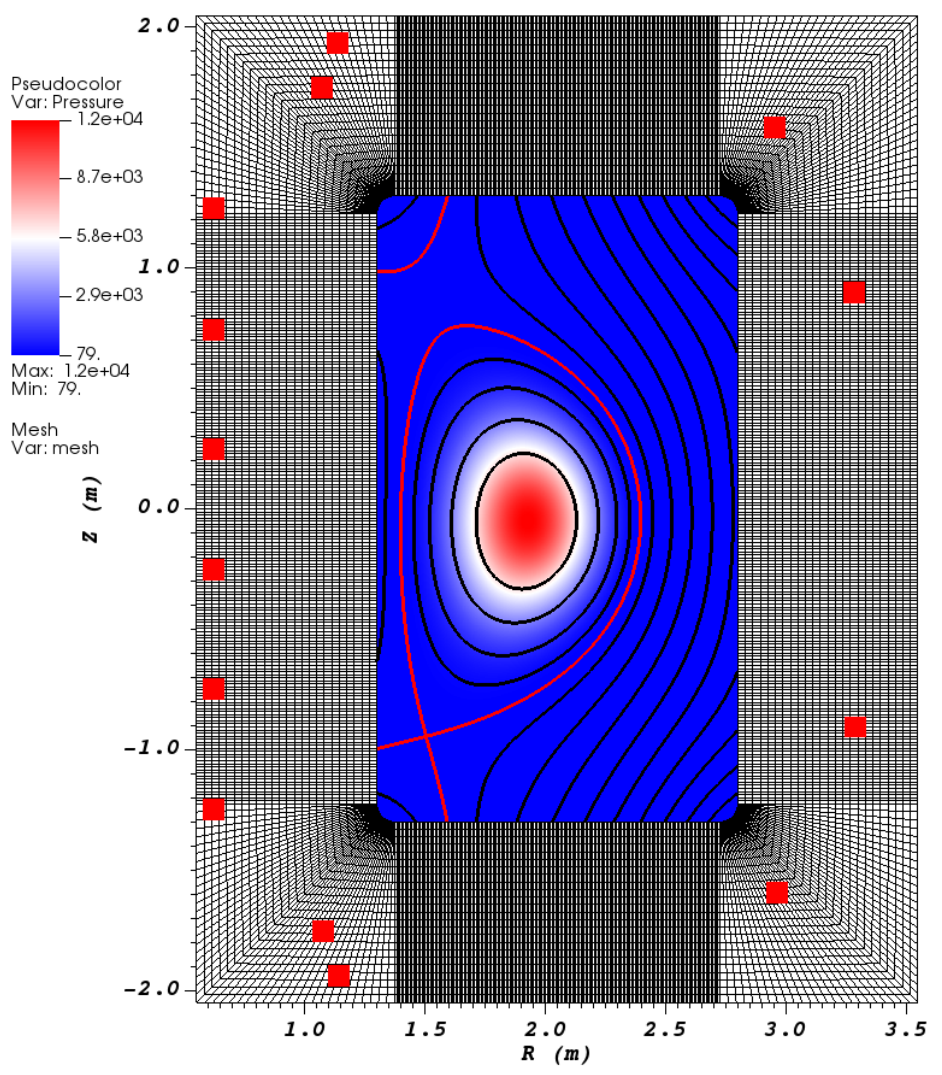}
  \caption{Restructed equilibrium using NIMEQ based on the equilibrium profiles of EAST experiment (presented in Fig.\ref{fig_VDE_rect:eq_profs}) and the poloidal field coil set of EAST device where the contour color denotes the pressure distribution, the dark contour lines denote the poloidal magnetic flux, the red line the separatrix, and the red rectangles stand for the locations of poloidal field coils.}
  \label{fig_VDE_rect:eq_psi_pres}
\end{figure}

The computational region is separated into two subdomains shown in Fig.\ref{fig_VDE_rect:eq_psi_pres}. The inner subdomain simulates the plasma activities with the visco-resistive MHD equations Eqs.(\ref{eqn:continue}-\ref{eqn:Ja_eval}). Eq.(\ref{eqn:Be_eval}) is advanced with a fixed magnetic diffusivity $\eta/\mu_{0}=100 \si{m^{2} \cdot s^{-1}}$ in the outer subdomain. The two subddomains are coupled using the thin-wall approximation
\begin{eqnarray}
\frac{\partial \left( \vec{B} \cdot \hat{n} \right)}{\partial t}=-\hat{n} \cdot \nabla \times \left[ \nu_{w} \hat{n} \times \Delta \vec{B} \right]
\end{eqnarray}
where $\hat{n}$ denotes the unit vector normal to the wall surface, $\nu_{w} \equiv \eta_{w}/\left( \mu_{0} \Delta_{w} \right)$ is defined as the ratio of the wall magnetic diffusivity to thickness, and $\Delta \vec{B}$ denotes the jump in magnetic field across the wall. The resistive wall lies along the boundary of inner subdomain. The term on the right side of Eq.(\ref{eqn:continue}) provides numerical smoothing to the particle number density field, and the coefficient value of $D_{n}=1.0$ used has negligible impact on the physically meaningful results~\cite{sovinec2019,nimrod}.

\begin{figure}[htb]
  \centering

	\begin{subfigure}[htb]{0.8\textwidth}
		\centering
		\includegraphics[width=1.0\textwidth]{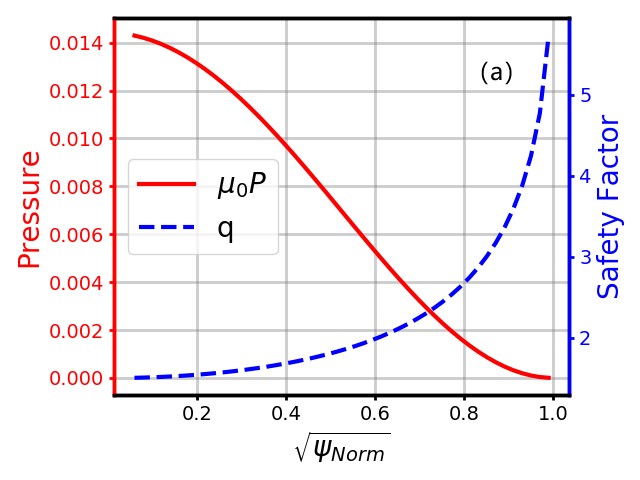}
		\label{fig_VDE_rect:eq_profs_pq}
	\end{subfigure}
	
    \begin{subfigure}[htb]{0.8\textwidth}
		\centering
		\includegraphics[width=1\textwidth]{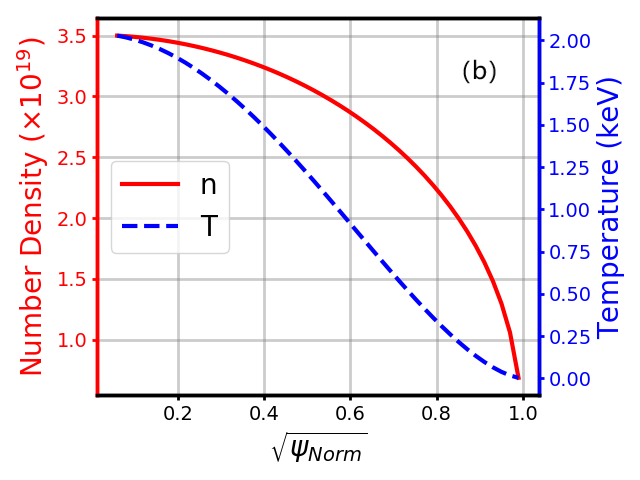}
		\label{fig_VDE_rect:eq_profs_nd_T}
	\end{subfigure}
  \caption{(a) Pressure and safety factor profiles of EAST experiment \#71230 at $48ms$ and (b) prescribed number density and temperature profiles of the equilibrium, where $\psi_{Norm}$ denotes the normalized poloidal magnetic flux.}
  \label{fig_VDE_rect:eq_profs}
\end{figure}

The initial equilibrium for simulations is reconstructed using NIMEQ~\cite{nimeq,nimeq_my} with the coil set of EAST device and the equilibrium profiles from experiment \#71230 at 48 ms (Fig.\ref{fig_VDE_rect:eq_profs})~\cite{ZAFAR_2021,DUAN2015727,Chen_2018}. The equilibrium particle number density profile is prescribed as $n=n_{0} \left( P/P_{0} \right)^{1/5}$. The key equilibrium parameters are listed in Table \ref{tab:eqs}.

\begin{table}[htb]
  \centering\small
  \caption{The key equilibrium parameters used in the simulation}
  \label{tab:eqs}
  \begin{tabular}{llll}
    \toprule
    Description  & Symbol   & Value   & Unit  \\ 
    \midrule
    Minor radius                     & $a$         & 0.476     & \si{m} \\
    Major radius                     & $R_{0}$     & 1.92      & \si{m} \\
    Elongation                       & $\kappa$    & 1.77      & 1 \\
    Triangularity                    & $\delta$    & 0.685     & 1 \\
    Toroidal magnetic field          & $B_{t0}$    & 1.73      & \si{T} \\ 
    Number density at magnetic axis  & $n_{0}$     & $3.5 \times 10^{19}$  & \si{m^{-3}} \\
    Temperature at magnetic axis     & $T_{0}$     & $1.0 \times 10^{3}$  & \si{eV} \\
    Equilibrium velocity             & $V_{0}$     & 0         & \si{m \cdot s^{-1}} \\
    Safety factor at magnetic axis   & $q_{0}$     & 1.5       & 1 \\ 
    $\beta$ at magnetic axis         & $\beta_{0}$ & 0.89\%    & 1 \\
    Alfv{\'e}n time                  & $\tau_{A}$  & $4.3 \times 10^{-7} \si{s}$   &  \si{s}  \\
    Lundquist number                 & $S$         & $5 \times 10^{5}$     &   1  \\
    Prandtl number                   & $Pr$        & 68       & 1 \\
    Parallel thermal conduction coefficent  & $\chi_{\parallel} $ & $ 1.0 \times 10^{6} $   & \si{m^{2}  \cdot s^{-1}} \\
    Perpendicular thermal conduction coefficent & $\chi_{\perp} $  & $1.0$    & \si{m^{2} \cdot s^{-1}} \\
    The ratio of the wall's magnetic diffusivity and thickness  & $\nu_{w}$  & $1.0 \times 10^{-3}$  & \si{m  \cdot s^{-1}} \\
    \bottomrule
  \end{tabular}
\end{table}

\section{Simulation results}
\label{results}

\subsection{Linear stability of initial equilibrium to non-axisymmetric perturbation}
\label{linear}

\begin{figure}[htb]
  \centering
  \includegraphics[width=1\textwidth]{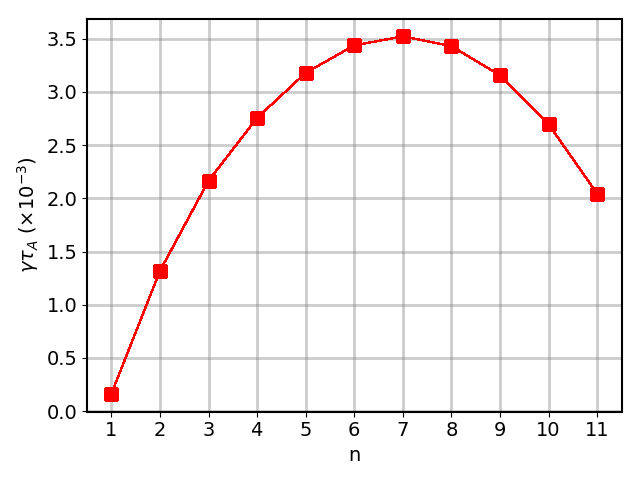}
  \caption{The normalized linear growth rate of the initial equilibrium as a function of the toroidal mode number $n$.}
  \label{linear_gamma}
\end{figure}

\begin{figure}[htb]
  \centering

	\begin{subfigure}[htb]{0.49\textwidth}
		\centering
		\includegraphics[width=1.0\textwidth]{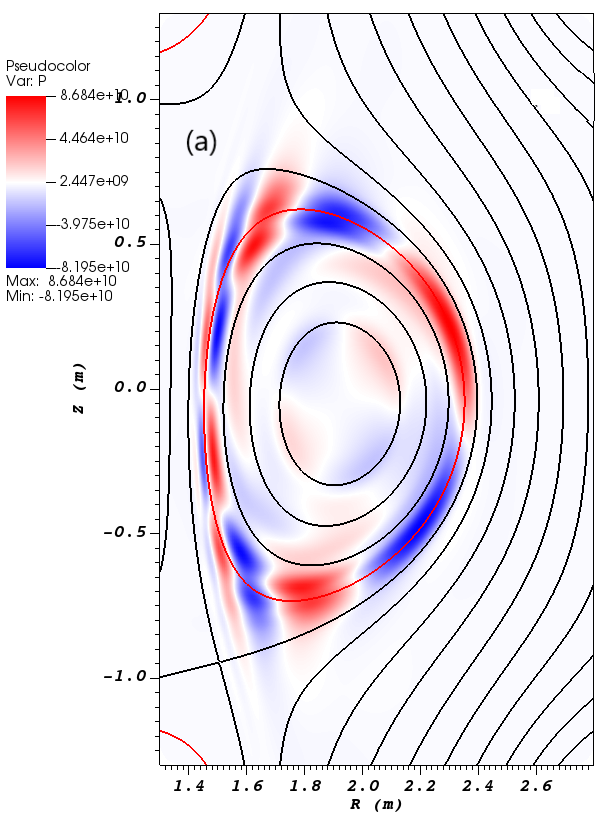}
		\label{linear_mode_structure_n1_pres}
	\end{subfigure}
    \begin{subfigure}[htb]{0.49\textwidth}
		\centering
		\includegraphics[width=1\textwidth]{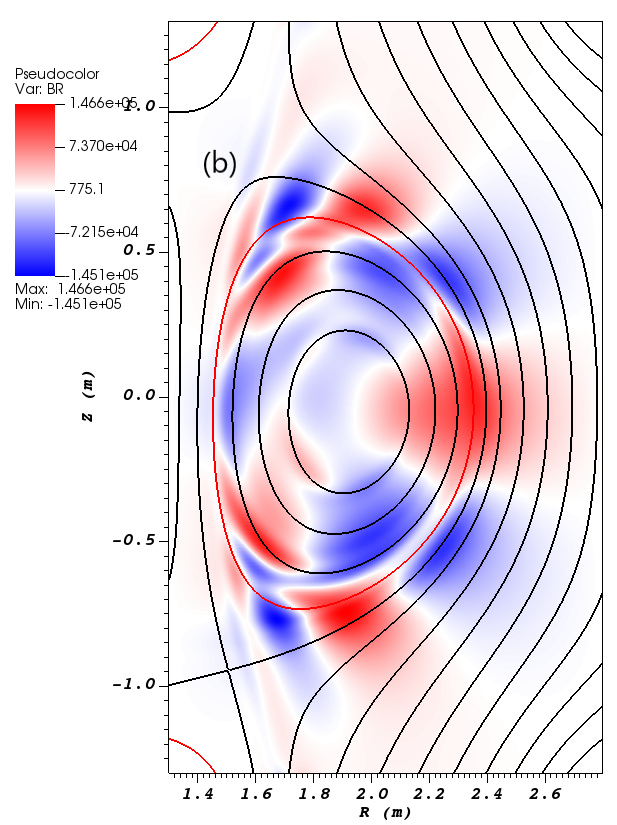}
		\label{linear_mode_structure_n1_BR}
	\end{subfigure}

\caption{Contours of the perturbed (a) pressure and (b) radial component of magnetic field from the linear $n=1$ mode (color) along with the poloidal magnetic flux (lines) of the initial equilibrium in the poloidal plane. The red line stands for the resonant surface where $q=4$.}
\label{linear_mode_structure_n1}
\end{figure}

To study the effects of the non-axisymmetric perturbations on the development of VDEs, the linear stability of the initial equilibrium to these perturbations are evaluated first here. From the linear NIMROD calculations, all asymmetric toroidal modes are unstable and the $n=7$ mode has the maximum growth rate (Fig.\ref{linear_gamma}). The $n=1$ mode behaves like an external-kink mode. The spatial structures of $n=1$ pressure and $B_{R}$ perturbations in poloidal plane are global and spread outside the separatrix (Fig.\ref{linear_mode_structure_n1}). The mode structures of all higher $n$ ($n \geq 1$) modes are located inside the separatrix, which indicates their internal mode nature. All mode structures are ballooning-like and appear around the same resonant surface $q=4$ close to separatrix  with different poloidal mode numbers. Thus, the $q=4$ resonant surface of the initial equilibrium is likely to be destroyed first as these instabilities develop into the nonlinear stage, which is consistent with the nonlinear simulation result.

\subsection{Nonlinear evolution of non-axisymmetric VDE}
\label{nonlinear}

\begin{figure}[htb]
  \centering

	\begin{subfigure}[htb]{0.8\textwidth}
		\centering
		\includegraphics[width=1.0\textwidth]{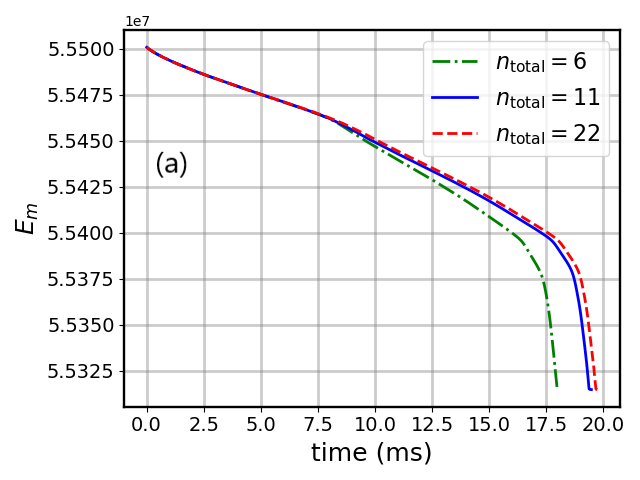}
		\label{3D_trunc_Em}
	\end{subfigure}
	
    \begin{subfigure}[htb]{0.8\textwidth}
		\centering
		\includegraphics[width=1\textwidth]{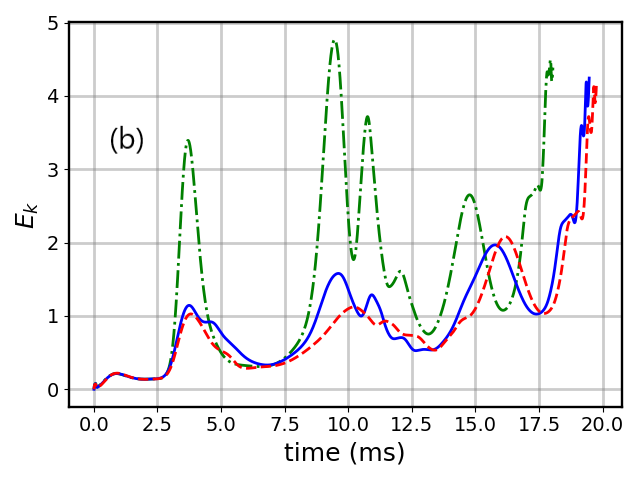}
		\label{3D_trunc_Ek}
	\end{subfigure}

\caption{The evolution of the magnetic energy and kinetic energy of the $n=0$ component over time during the nonlinear simulation of a non-axisymmetric VDE with various total numbers of toroidal components $n_{\rm{total}}$.}
\label{fig_VDE_rect:3D_trunc}
\end{figure}

\begin{figure}[htb]
  \centering
  \includegraphics[width=1\textwidth]{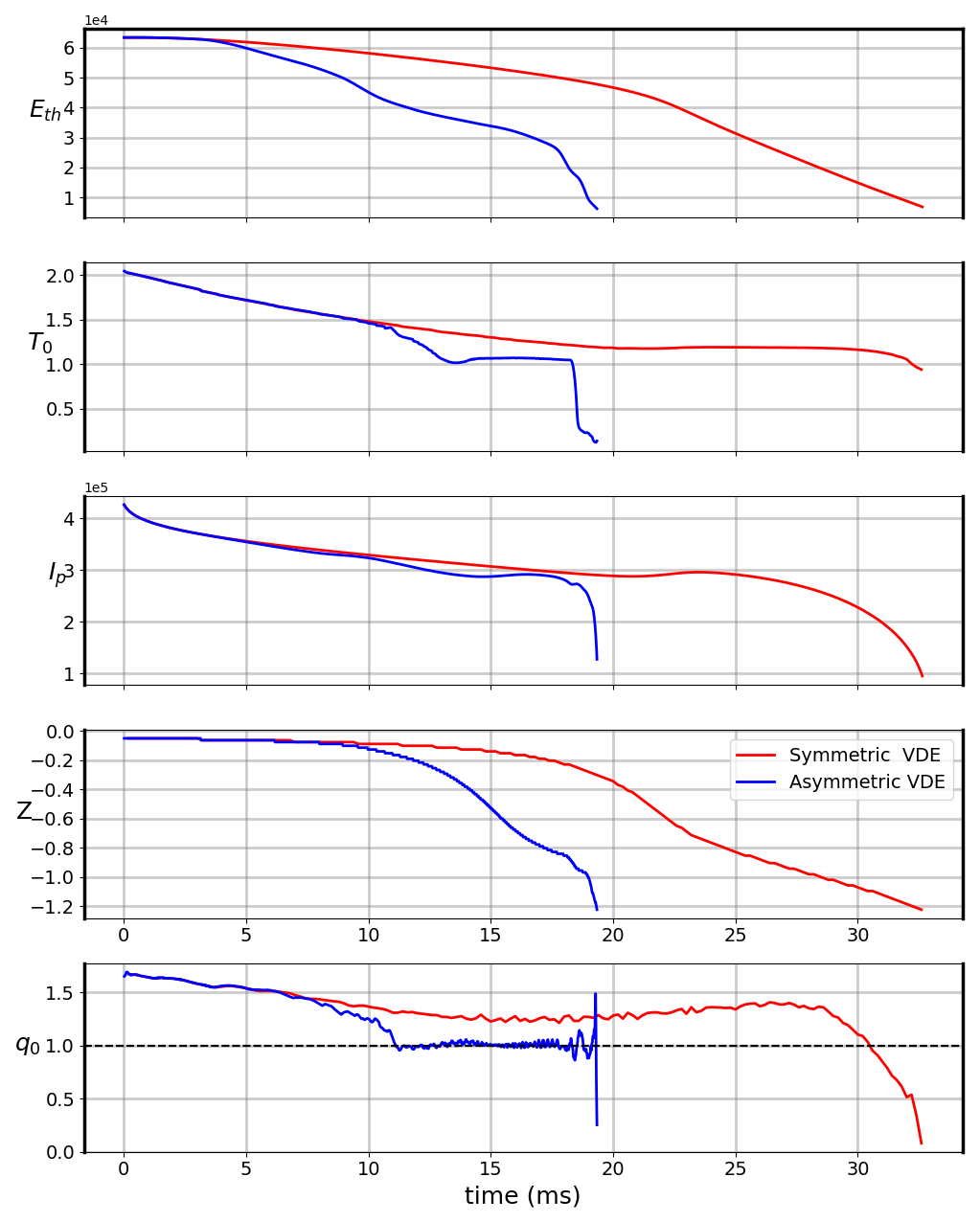}
  \caption{The evolution of the total internal energy $E_{th}$, the temperature at magnetic axis $T_{0}$, the plasma current $I_{p}$, the vertical coordinate of magnetic axis $Z$, and the safety factor at magnetic axis $q_{0}$ over time during axisymmetric VDE (red line) and non-axisymmetric VDE (blue line), where the internal energy is defined as $E_{\rm{th}}=\frac{k_{b} \left( T_{eq} + \tilde{T} \right) \left( n_{eq} + \tilde{n} \right)}{\gamma-1}$.}
  \label{fig_VDE_rect:3D_Eth_I_q0}
\end{figure}

\begin{figure}[htb]
  \centering
		\includegraphics[width=1\textwidth]{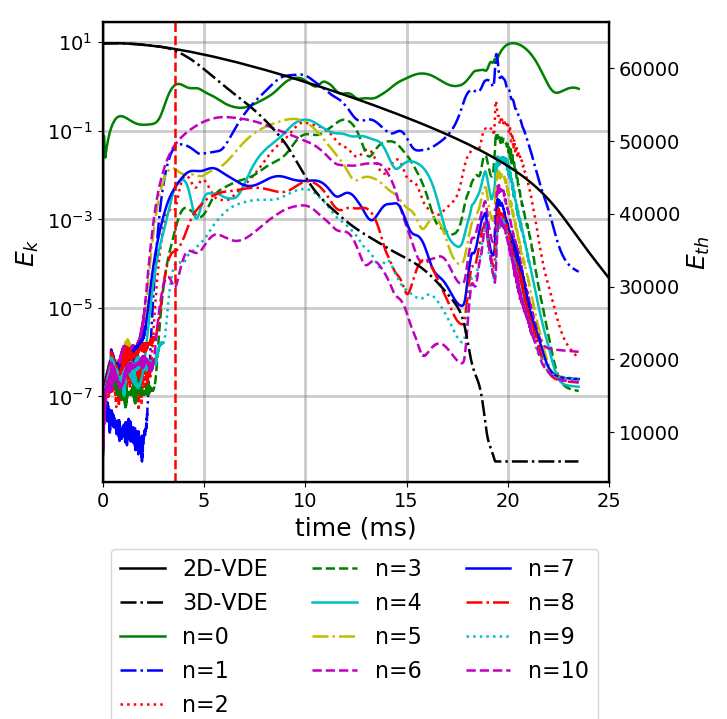}
  \caption{Time evolution of the kinetic energy ($E_{k}$) of various toroidal components during a non-axisymmetric VDE (color lines), and the total internal energy $E_{\rm{th}}$ during an axisymmetric VDE (solid dark line) and the non-axisymmetric VDE (dash-dot dark line). The red dashed vertical line stands for $t=3.6ms$.}
  \label{fig_VDE_rect:3D_Ek}
\end{figure}

\begin{figure}[htb]
  \centering
		\includegraphics[width=1\textwidth]{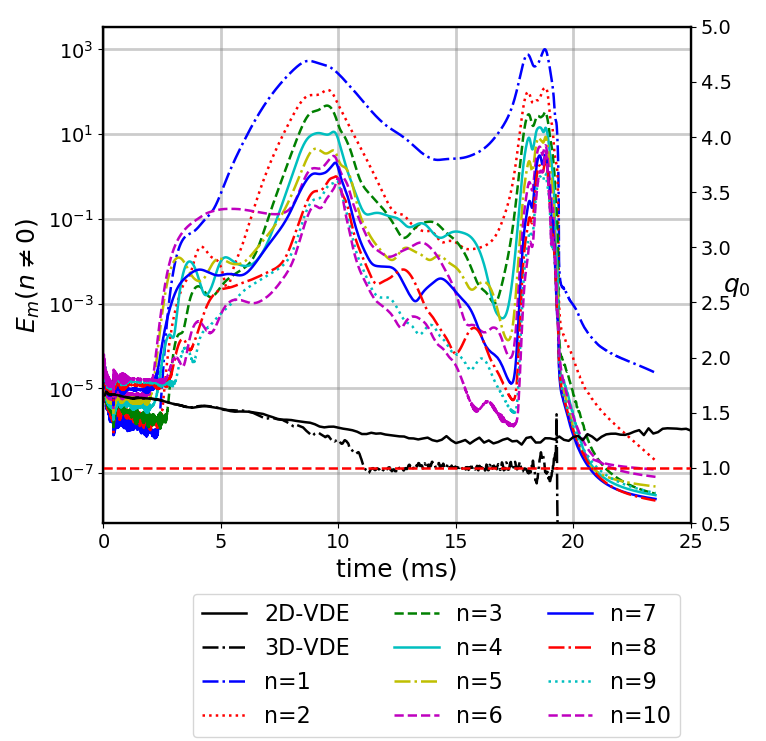}
  \caption{Time evolution of the magnetic energy ($E_{m}$) of various non-axisymmetric toroidal components ($n \neq 0$) during a non-axisymmetric VDE  (color lines), and the $q_{0}$ during an axisymmetric VDE (solid dark line) and the non-axisymmetric VDE (dash-dot dark line). The red dashed horizontal line stands for $q_{0}=1.0$.}
  \label{fig_VDE_rect:3D_Em}
\end{figure}

\begin{figure}[htb]
  \centering
  \includegraphics[width=1\textwidth]{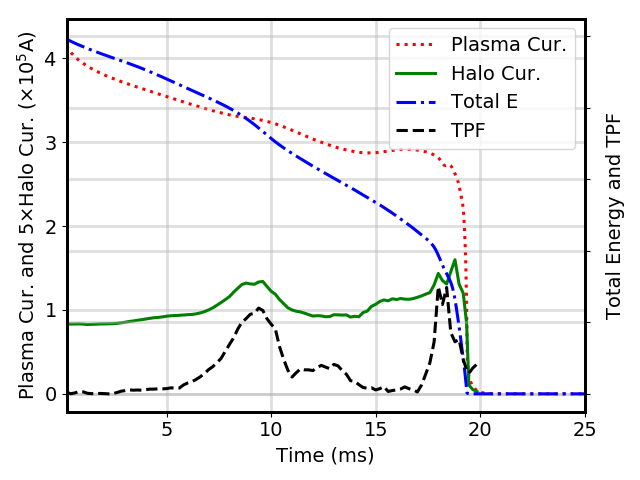}
  \caption{Time evolution of plasma current, halo current, total  energy (including both perturbation and equilibrium), and TPF during a non-axisymmetric VDE.}
  \label{fig_VDE_rect:halo_TPF}
\end{figure}

For $\chi_{\parallel}/\chi_{\perp} = 1.0 \times 10^{6}$, the inclusion of toroidal mode numbers from $0$ to $10$ ensures numerical convergence (Fig.\ref{fig_VDE_rect:3D_trunc}). The complete thermal quench (TQ) and current quench (CQ) are observed clearly along with both axisymmetric and non-axisymmetric VDEs (Fig.\ref{fig_VDE_rect:3D_Eth_I_q0}). The behaviors of plasma are similar during the startup stage. The difference in total internal energy between the axisymmetric and the non-axisymmetric VDEs shows up from $t=3.6 \si{ms}$, which is the moment that all asymmetric toroidal modes reach saturation (Fig.\ref{fig_VDE_rect:3D_Ek}). The evolution of perturbation energy indicates that there are two disruption events in the non-axisymmetric VDE process (Figs.\ref{fig_VDE_rect:3D_Ek}-\ref{fig_VDE_rect:3D_Em}). The minor disruption occurs around $t=10 \si{ms}$, and the major disruption leads to the termination of plasma (Fig.\ref{fig_VDE_rect:halo_TPF}). The non-axisymmetric perturbations in nonlinear simulation tend to accelerate the vertical motion of plasma, which suggests the direct effects of three dimensional physics. 

\begin{figure}[!htb]
  \centering

	\begin{subfigure}[htb]{0.49\textwidth}
		\centering
		\includegraphics[width=1\textwidth]{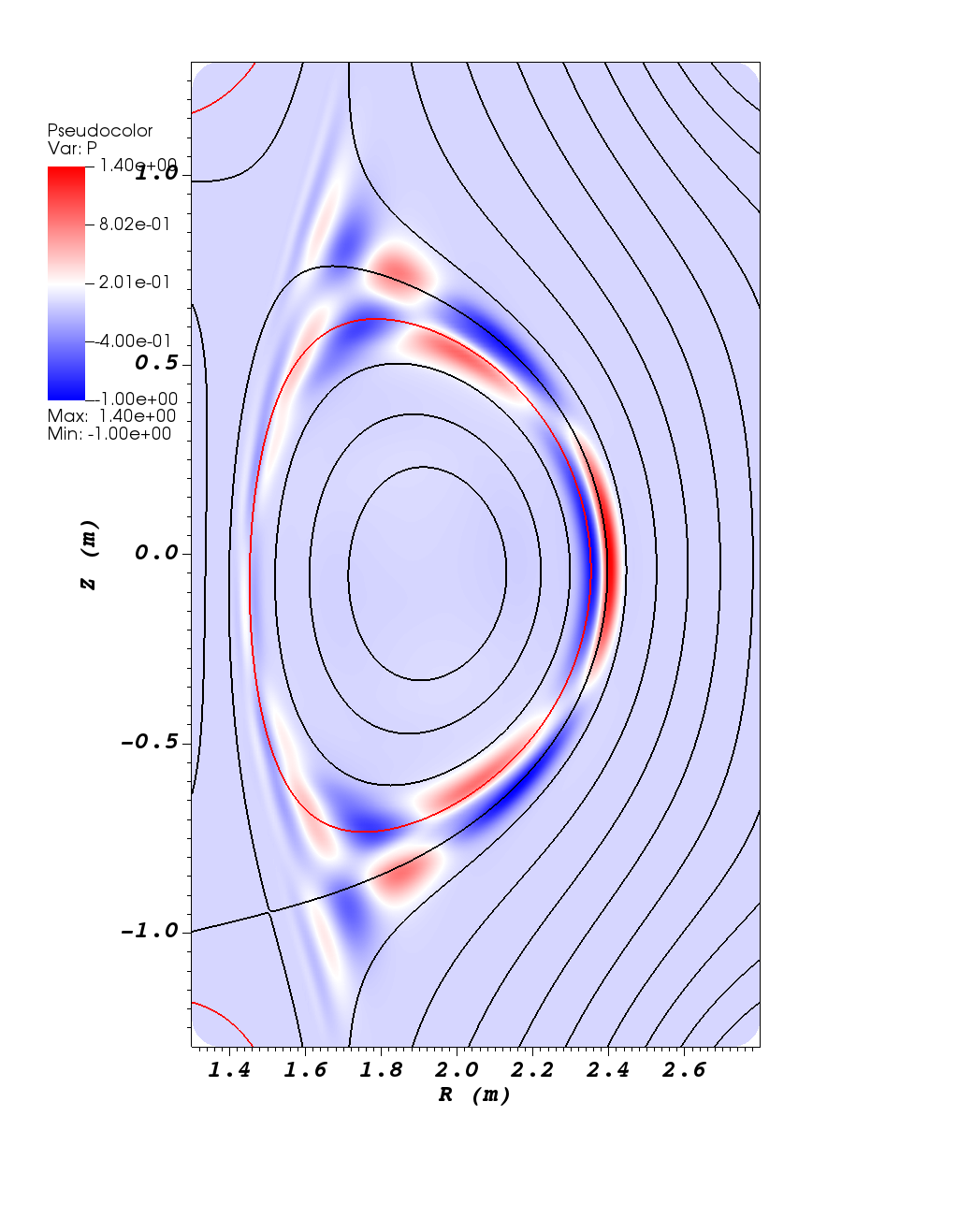}
		\caption{$n=1$ at $t=2.5\si{ms}$}
		\label{fig_VDE_rect:quad_con_n1_250}
	\end{subfigure}
	\begin{subfigure}[htb]{0.49\textwidth}
		\centering
		\includegraphics[width=1\textwidth]{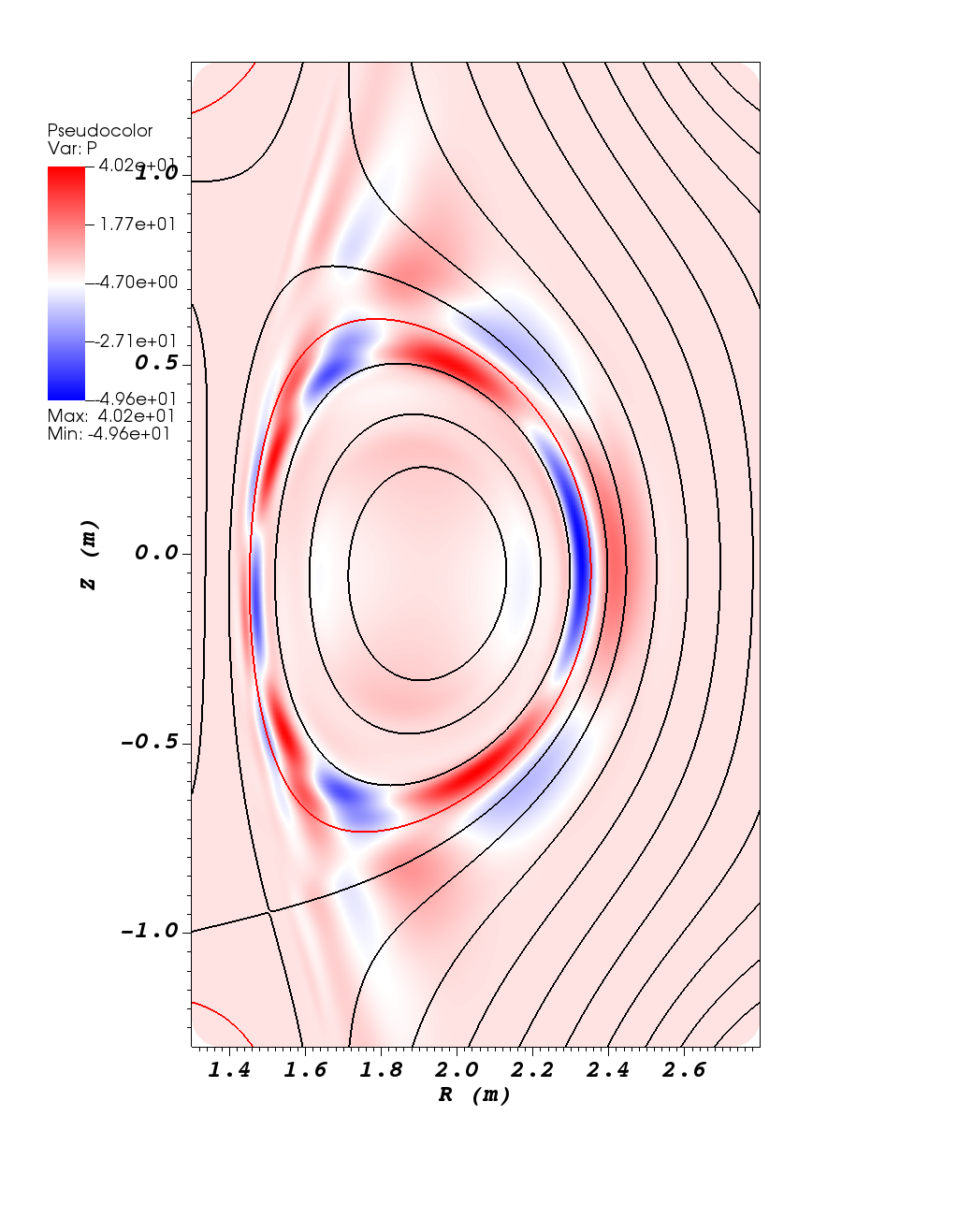}
		\caption{$n=1$ at $t=3.6\si{ms}$}
		\label{fig_VDE_rect:quad_con_n1_360}
	\end{subfigure}
	
	\begin{subfigure}[htb]{0.49\textwidth}
		\centering
		\includegraphics[width=1\textwidth]{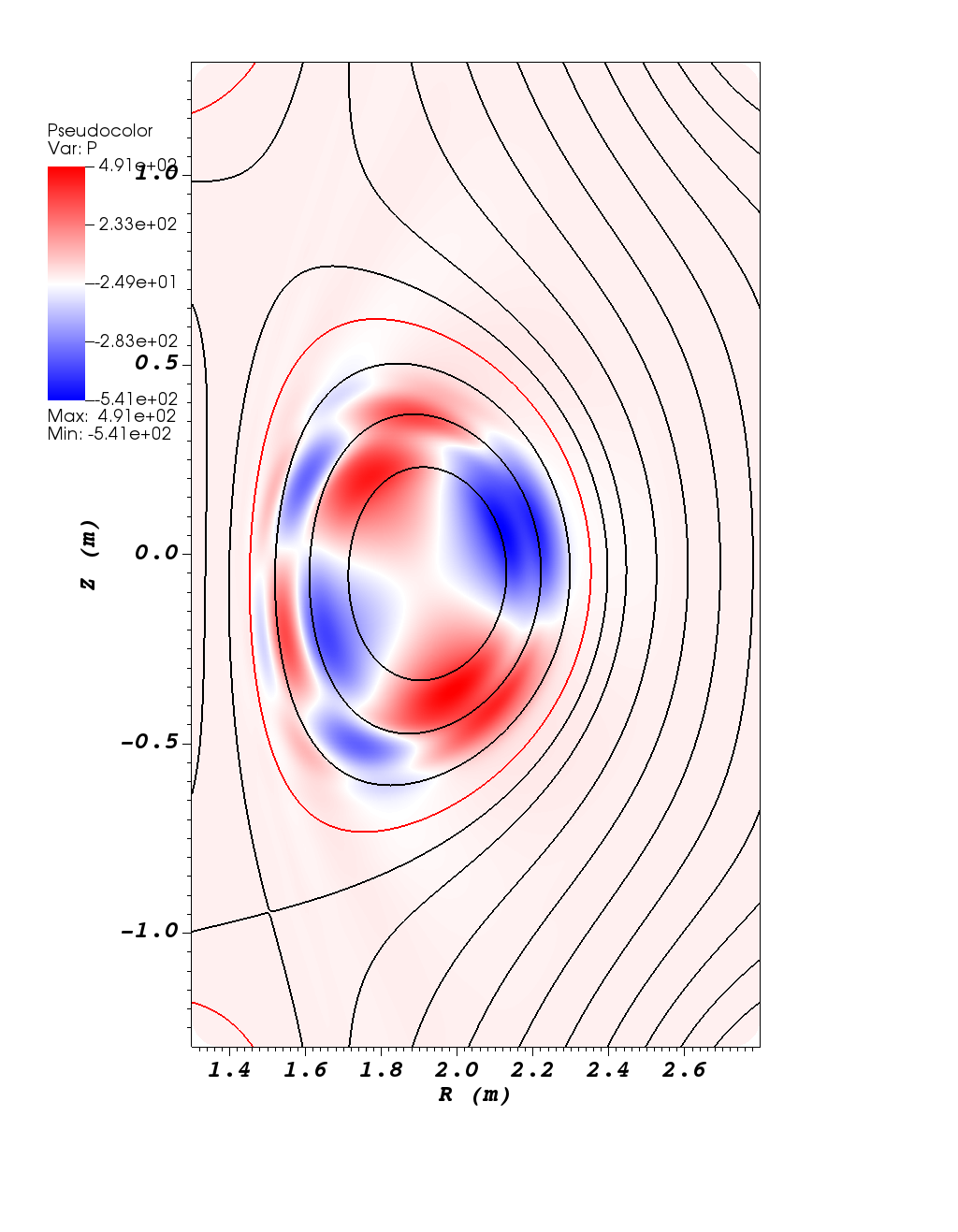}
		\caption{$n=1$ at $t=7.2\si{ms}$}
		\label{fig_VDE_rect:quad_con_n1_720}
	\end{subfigure}
	\begin{subfigure}[htb]{0.49\textwidth}
		\centering
		\includegraphics[width=1\textwidth]{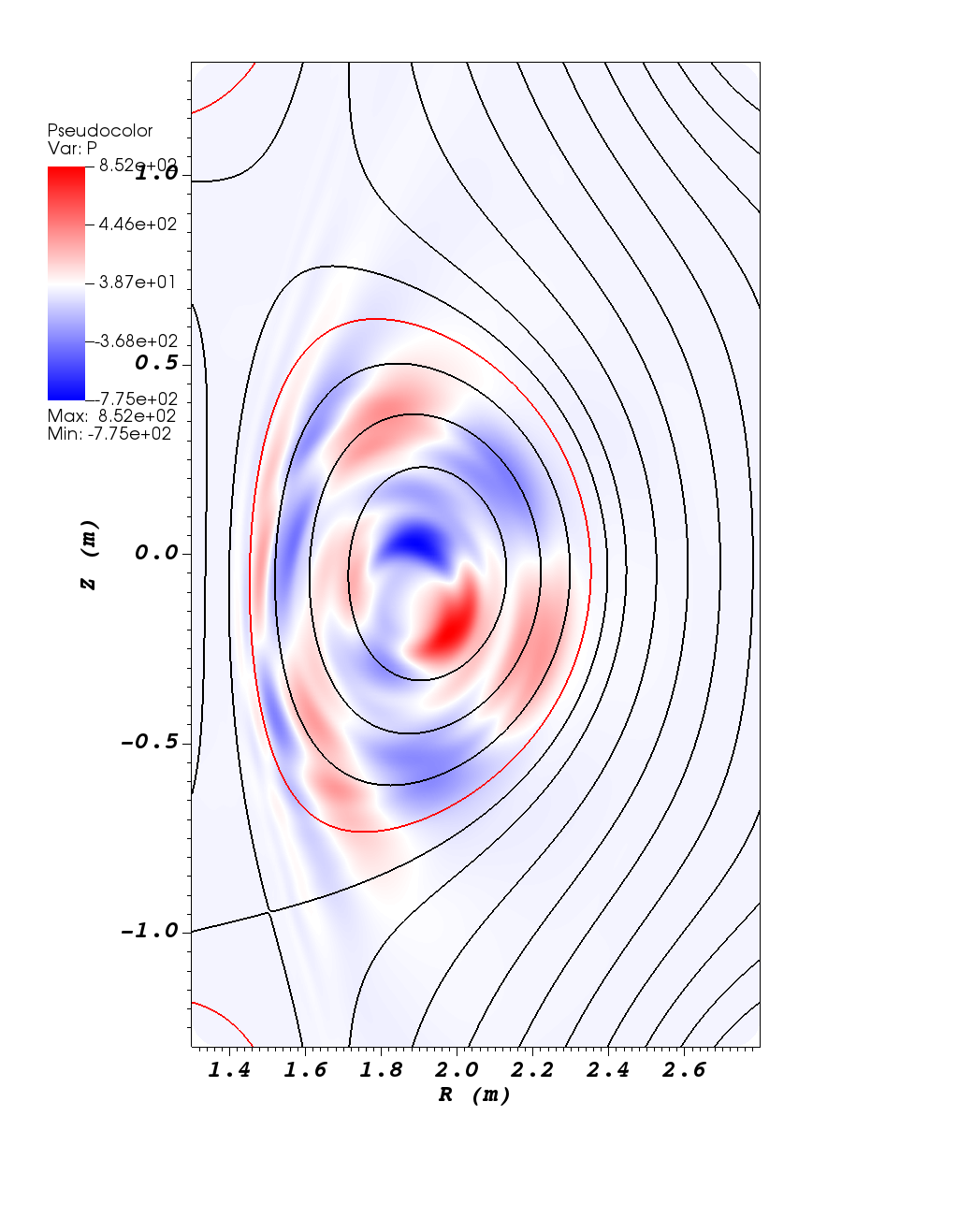}
		\caption{$n=1$ at $t=10.0\si{ms}$}
		\label{fig_VDE_rect:quad_con_n1_1000}
	\end{subfigure}

\caption{Contours of pressure perturbation (color) of the $n=1$ component in the poloidal plane at (a) $t=2.5\si{ms}$, (b) $t=3.6\si{ms}$, (c) $t=7.2\si{ms}$, and (d) $t=10.0\si{ms}$, during a non-axisymmetric VDE and the poloidal magnetic flux of the initial equilibrium (line). The red line stands for the resonant surface where $q=4$.}
\label{fig_VDE_rect:quad_con_n1}

\end{figure}

\begin{figure}[htb]
  \centering

	\begin{subfigure}[htb]{0.49\textwidth}
		\centering
		\includegraphics[width=1.0\textwidth]{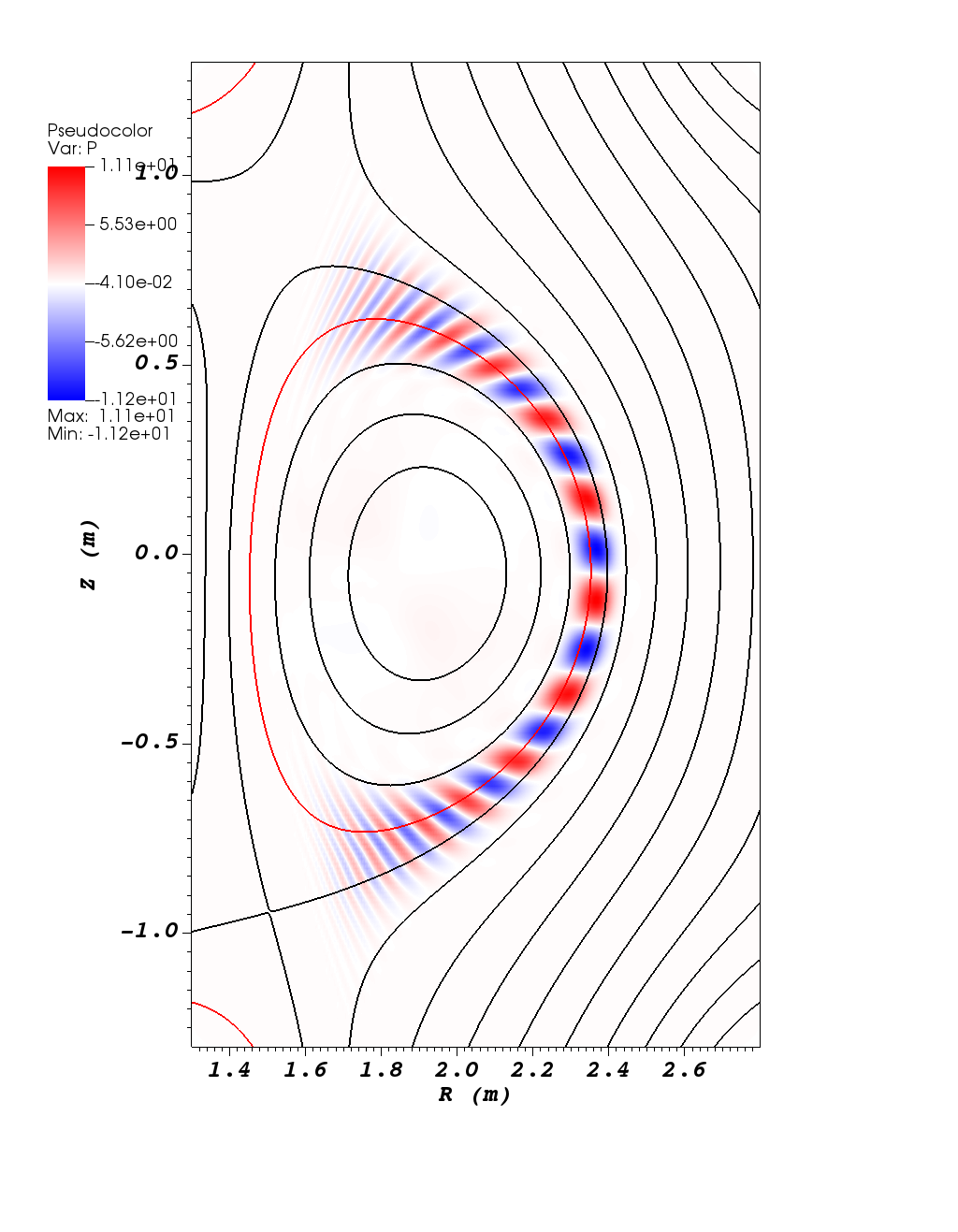}
		\caption{$n=6$ at $t=2.5\si{ms}$.}
		\label{fig_VDE_rect:quad_con_n6_250}
	\end{subfigure}\begin{subfigure}[htb]{0.49\textwidth}
		\centering
		\includegraphics[width=1\textwidth]{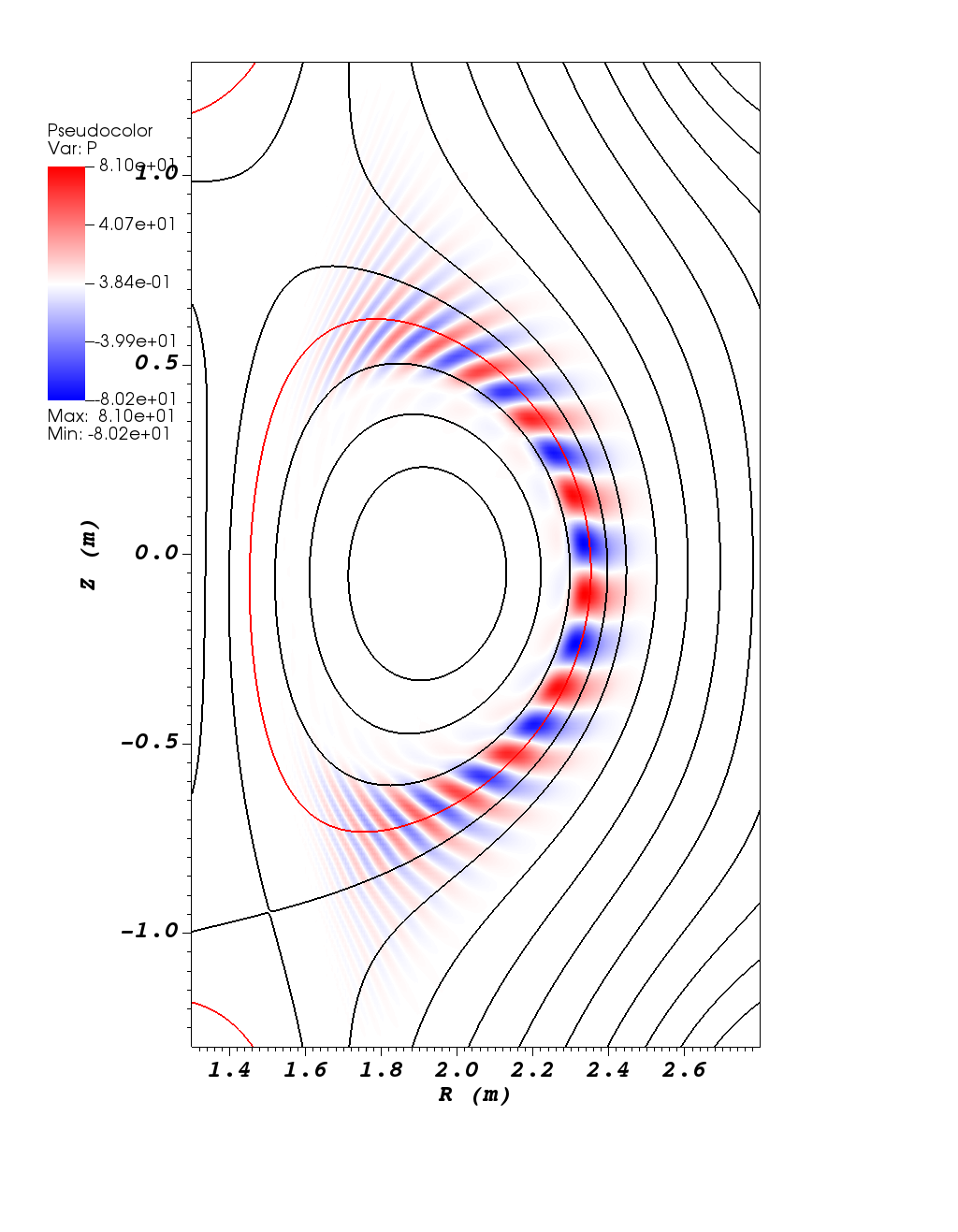}
		\caption{$n=6$ at $t=3.6\si{ms}$.}
		\label{fig_VDE_rect:quad_con_n6_360}
	\end{subfigure}

	\begin{subfigure}[htb]{0.49\textwidth}
		\centering
		\includegraphics[width=1.0\textwidth]{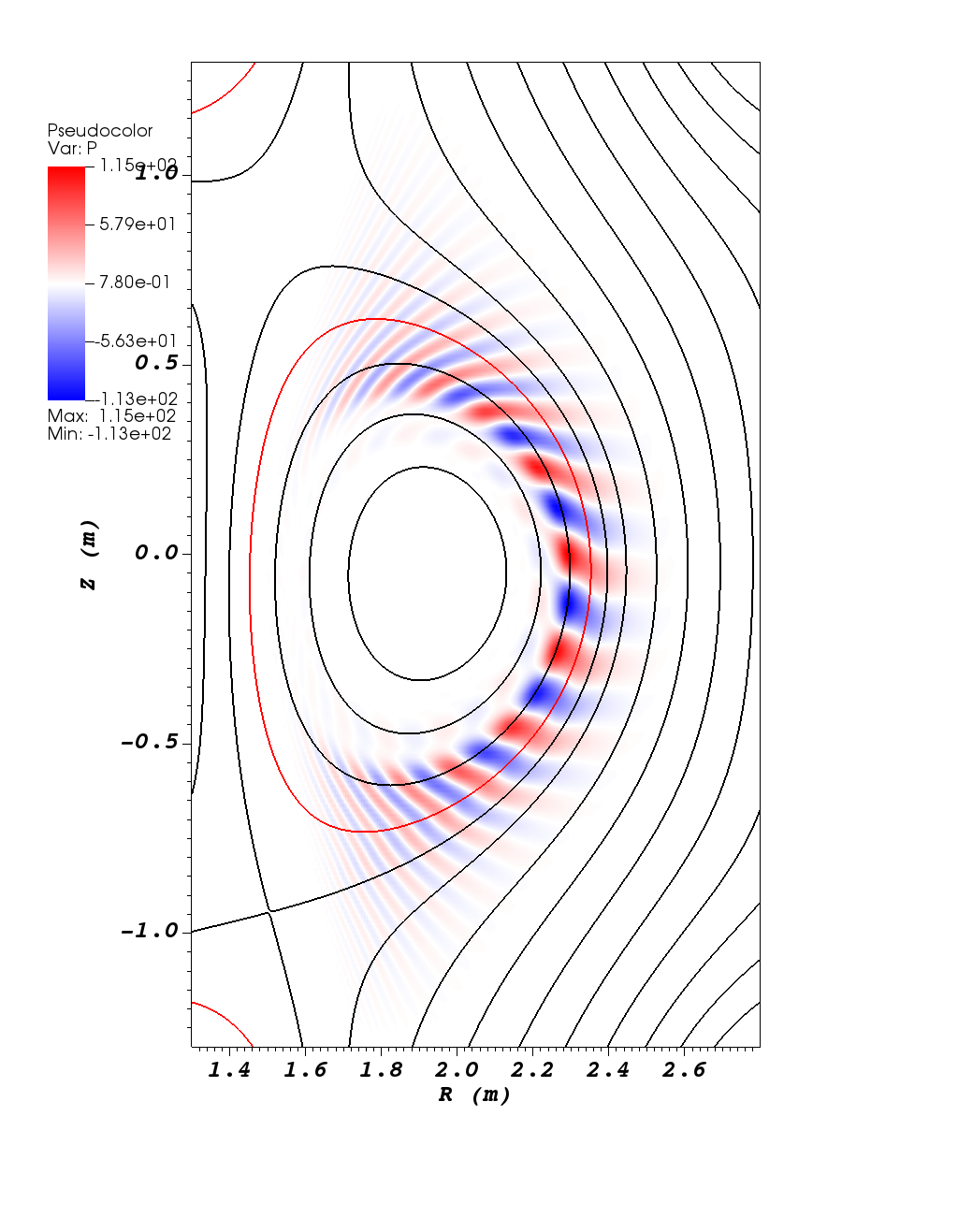}
		\caption{$n=6$ at $t=7.2\si{ms}$.}
		\label{fig_VDE_rect:quad_con_n6_720}
	\end{subfigure}\begin{subfigure}[htb]{0.49\textwidth}
		\centering
		\includegraphics[width=1\textwidth]{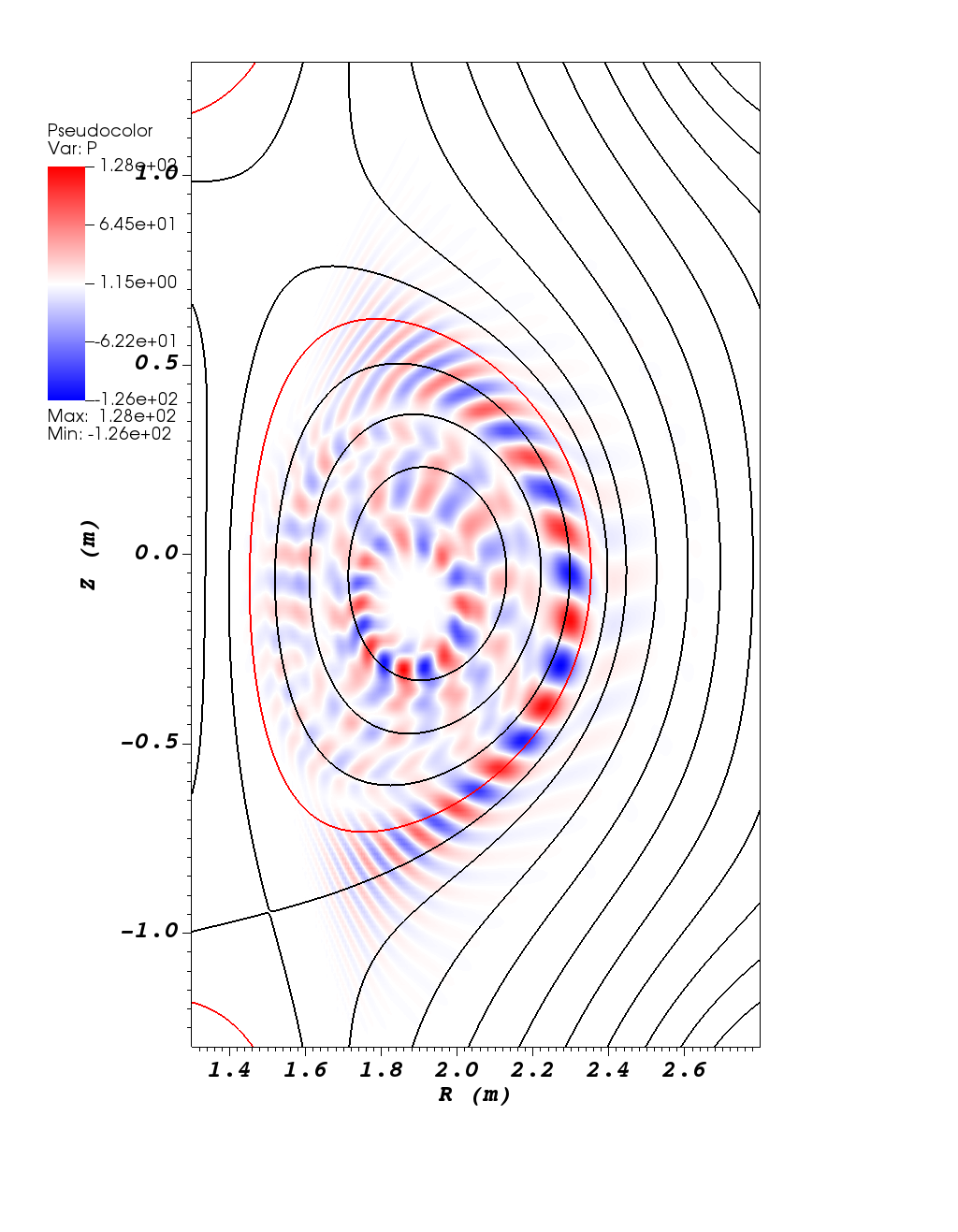}
		\caption{$n=6$ at $t=10.0\si{ms}$.}
		\label{fig_VDE_rect:quad_con_n6_1000}
	\end{subfigure}

\caption{Contours of pressure perturbation (color) of the $n=6$ component in the poloidal plane at (a) $t=2.5\si{ms}$, (b) $t=3.6\si{ms}$, (c) $t=7.2\si{ms}$, and (d) $t=10.0\si{ms}$, during a non-axisymmetric VDE and the poloidal magnetic flux of the initial equilibrium (line). The red line stands for the resonant surface where $q=4$.}
\label{fig_VDE_rect:quad_con_n6}
\end{figure}

\begin{figure}[htb]
  \centering

	\begin{subfigure}[htb]{0.49\textwidth}
		\centering
		\includegraphics[width=0.9\textwidth]{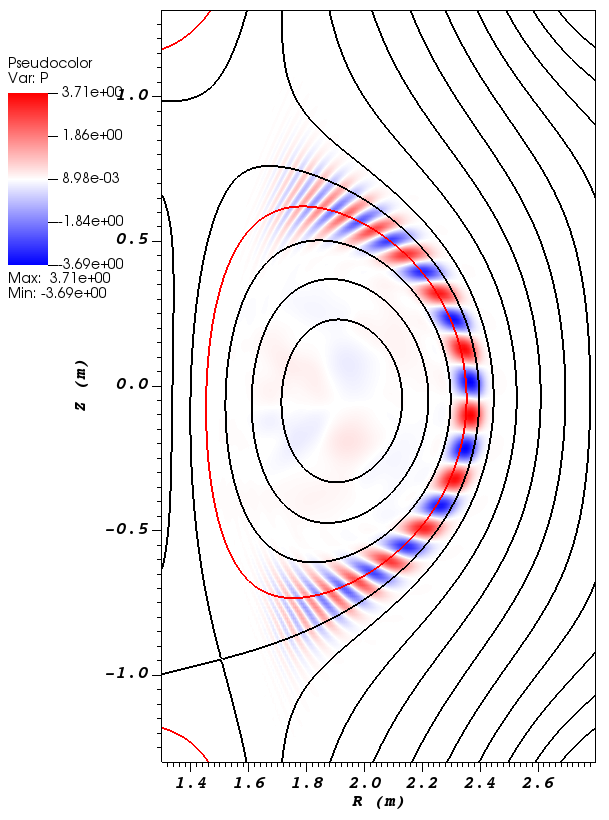}
		\caption{$n=7$ at $t=2.5\si{ms}$.}
		\label{fig_VDE_rect:quad_con_n7_250}
	\end{subfigure}\begin{subfigure}[htb]{0.49\textwidth}
		\centering
		\includegraphics[width=0.9\textwidth]{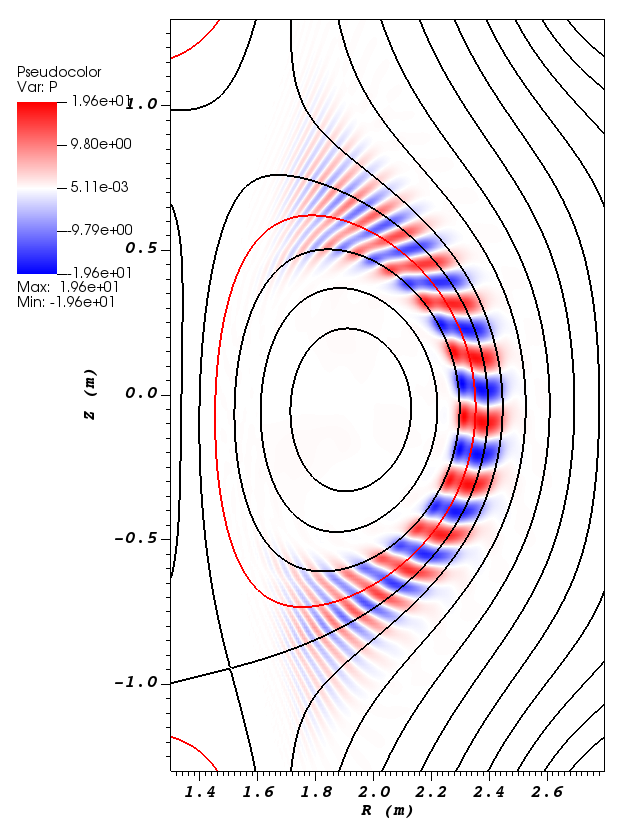}
		\caption{$n=7$ at $t=4.0\si{ms}$.}
		\label{fig_VDE_rect:quad_con_n7_400}
	\end{subfigure}

	\begin{subfigure}[htb]{0.49\textwidth}
		\centering
		\includegraphics[width=0.9\textwidth]{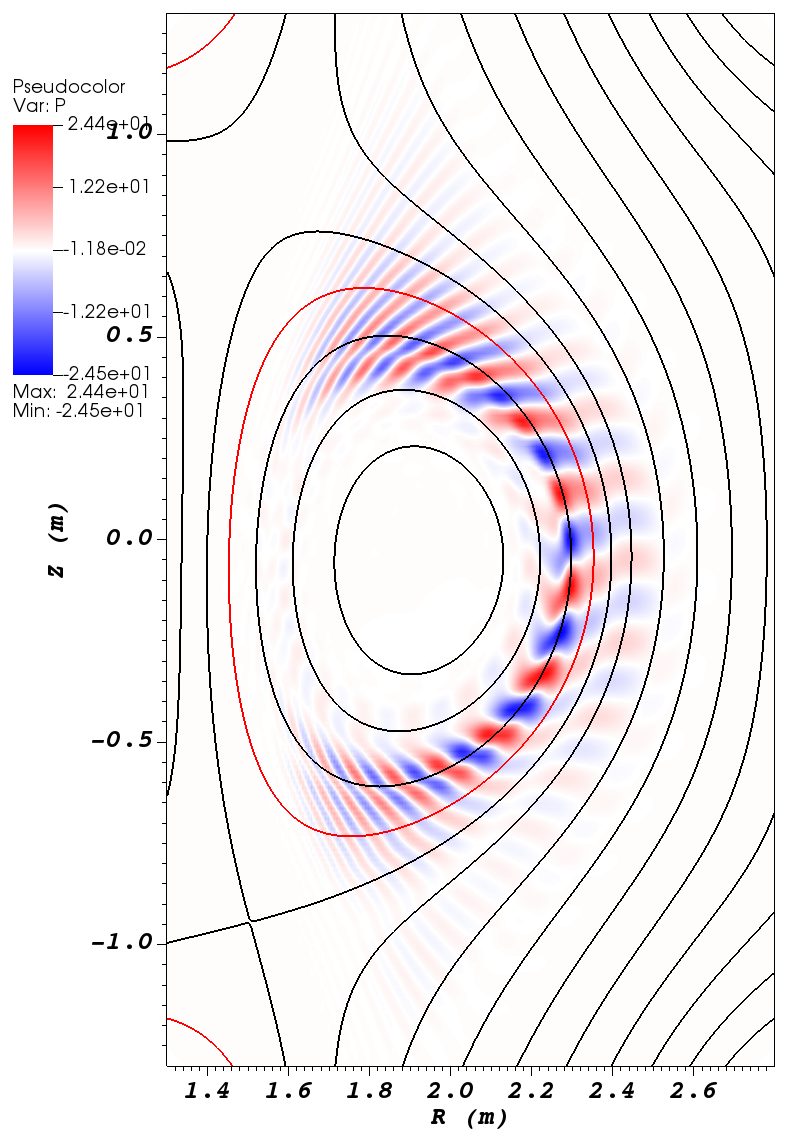}
		\caption{$n=7$ at $t=7.2\si{ms}$.}
		\label{fig_VDE_rect:quad_con_n7_720}
	\end{subfigure}\begin{subfigure}[htb]{0.49\textwidth}
		\centering
		\includegraphics[width=0.9\textwidth]{EAST_VDE_quad_con_lphi5_kpll1e6_00720_n7_v_flux_w_q4_20210529.png}
		\caption{$n=7$ at $t=10.0\si{ms}$.}
		\label{fig_VDE_rect:quad_con_n7_1000}
	\end{subfigure}

\caption{Contours of pressure perturbation (color) of the $n=7$ component in the poloidal plane at (a) $t=2.5\si{ms}$, (b) $t=4.0\si{ms}$, (c) $t=7.2\si{ms}$, and (d) $t=10.0\si{ms}$, during a non-axisymmetric VDE and the poloidal magnetic flux of the initial equilibrium (line). The red line stands for the resonant surface where $q=4$.}
\label{fig_VDE_rect:quad_con_n7}
\end{figure}

For the non-axisymmetric VDE, although the $n=7$ mode has the largest linear growth among all the non-axisymmetric toroidal components ($n \geq 1$) (Fig.\ref{linear_gamma}), the saturated energy levels of $n=1$ and $n=6$ modes are dominant, which are close to that of the $n=0$ component including the equilibrium contribution (Figs.\ref{fig_VDE_rect:3D_Ek}-\ref{fig_VDE_rect:3D_Em}), and thus can substantially affect the shape and motion of plasma. As the dominant non-axisymmetric component during the entire VDE process, the $n=1$ mode can reconfigure the magnetic flux surface from the boundary of plasma to the core region (Fig.\ref{fig_VDE_rect:quad_con_n1}). The mode structures of $n=6$ and $n=7$ components display finger filement structures around the $q=4$ resonant surface, which extend beyond the separatrix (Fis.\ref{fig_VDE_rect:quad_con_n6}-\ref{fig_VDE_rect:quad_con_n7}), even though the equilibrium has no edge pressure pedestal.

\begin{figure}[!htb]
  \centering

	\begin{subfigure}[htb]{0.9\textwidth}
		\centering
		\includegraphics[width=1\textwidth]{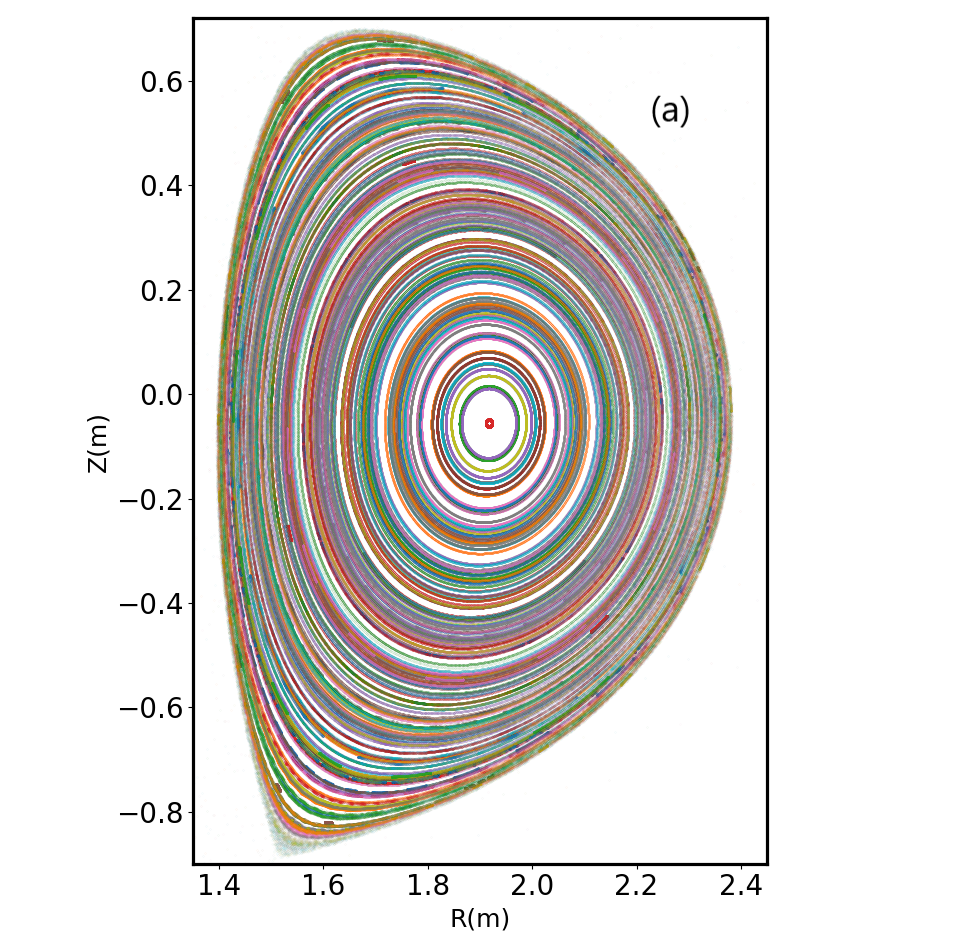}
		\label{fig_VDE_rect:quad_poincare_250_all}
	\end{subfigure}
	
    \begin{subfigure}[htb]{0.9\textwidth}
		\centering
		\includegraphics[width=1\textwidth]{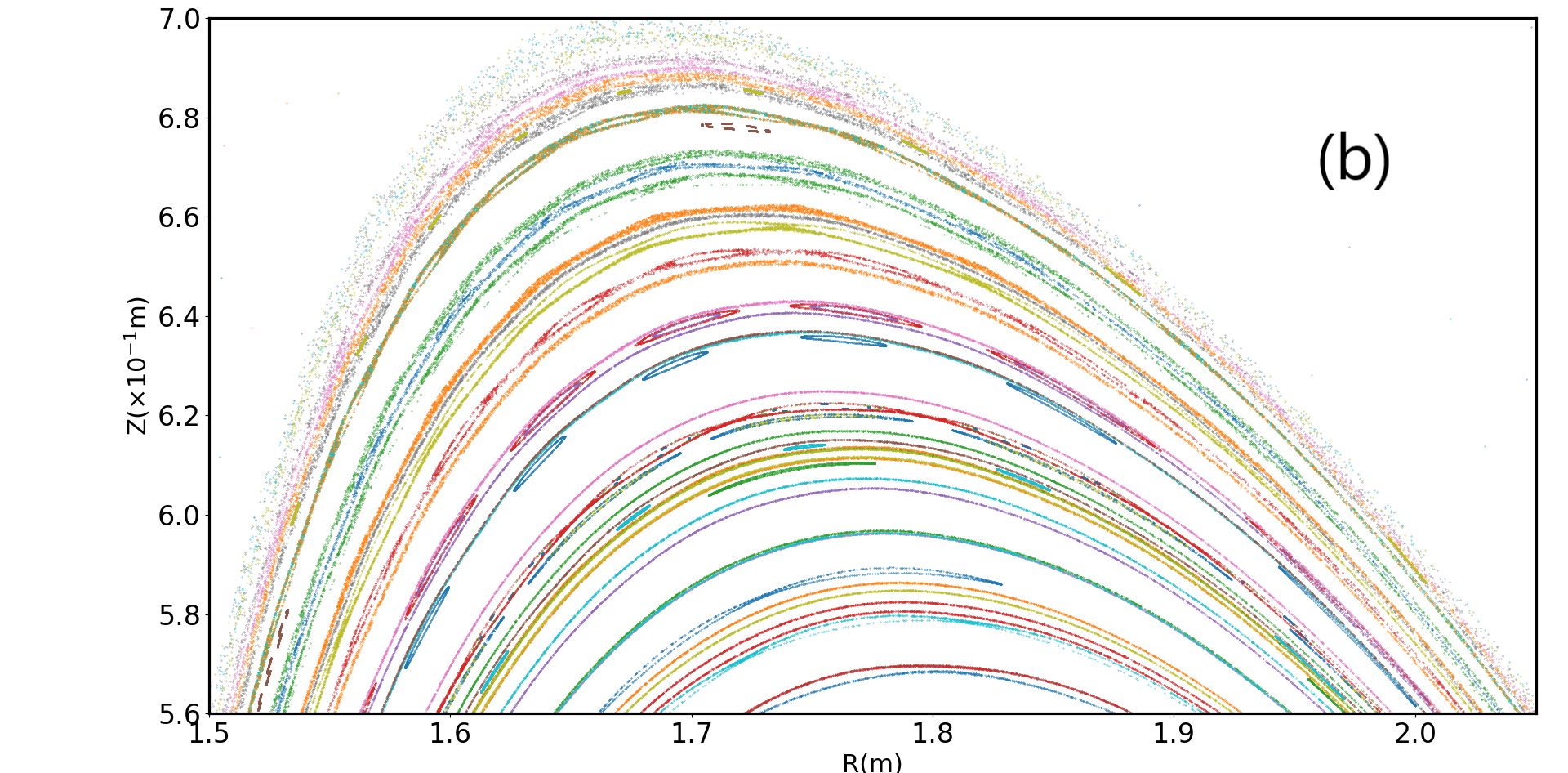}
		\label{fig_VDE_rect:quad_poincare_250_zoom}
	\end{subfigure}

\caption{The Poincar$\acute{\rm{e}}$ plot at $t=2.5\si{ms}$ in the poloidal plane during a non-axisymmetric VDE, where figure (b) is the zoomed in part of figure (a).}
\label{fig_VDE_rect:quad_poincare_250}

\end{figure}
\begin{figure}[!htb]
  \centering

	\begin{subfigure}[htb]{0.49\textwidth}
		\centering
		\includegraphics[width=1\textwidth]{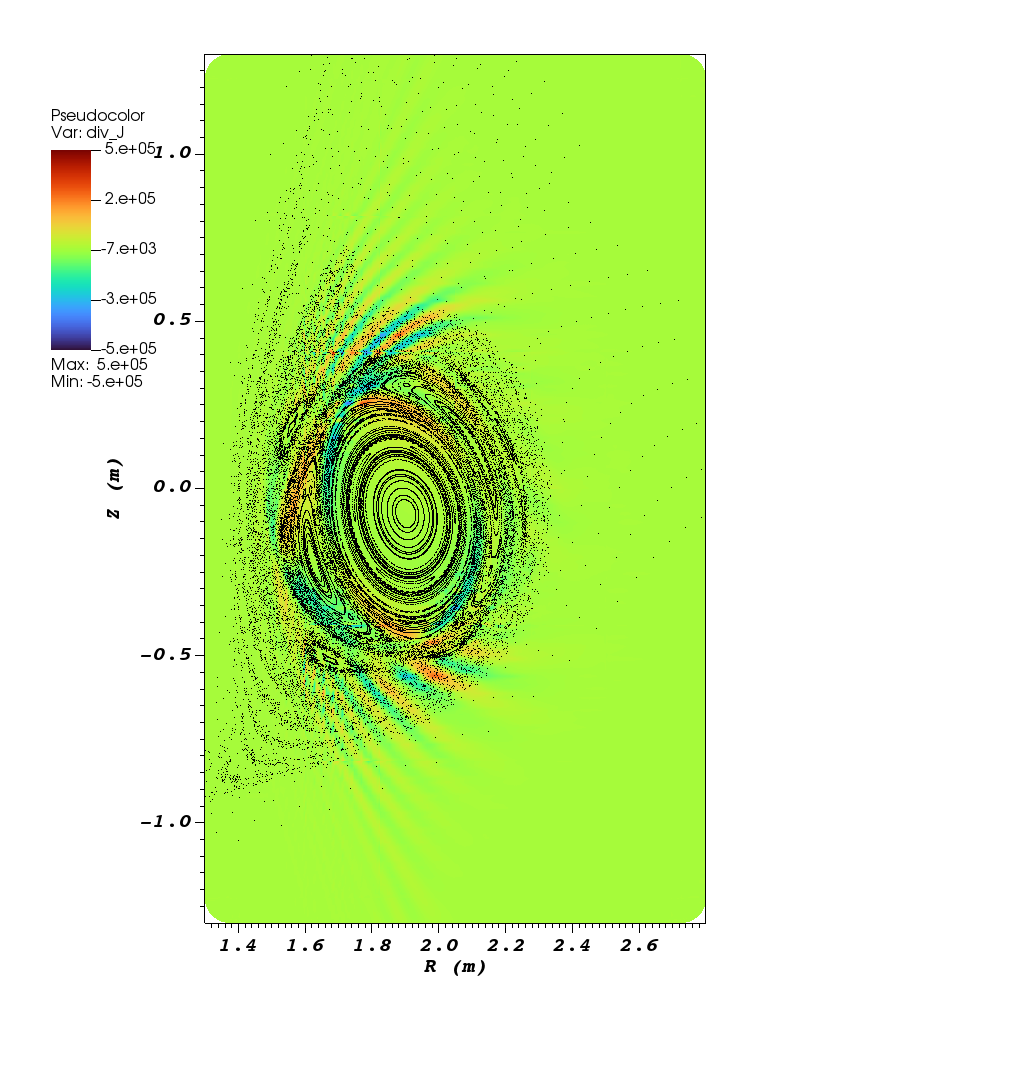}
		\caption{$t=7.2\si{ms}$}
		\label{fig_VDE_rect:div_J_poi_720}
	\end{subfigure}
    \begin{subfigure}[htb]{0.49\textwidth}
		\centering
		\includegraphics[width=1\textwidth]{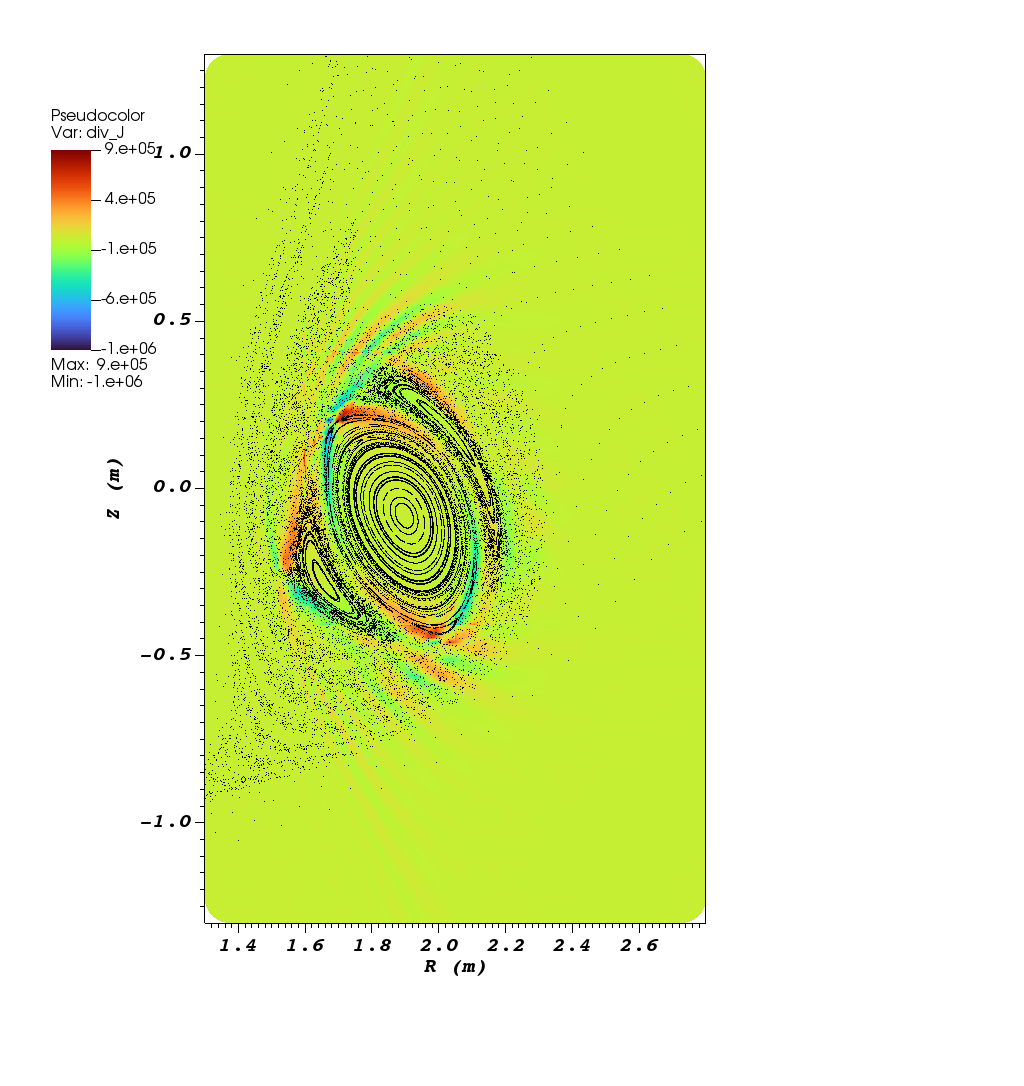}
		\caption{$t=8.0\si{ms}$}
		\label{fig_VDE_rect:div_J_poi_800}
	\end{subfigure}
	
	\begin{subfigure}[htb]{0.49\textwidth}
		\centering
		\includegraphics[width=1\textwidth]{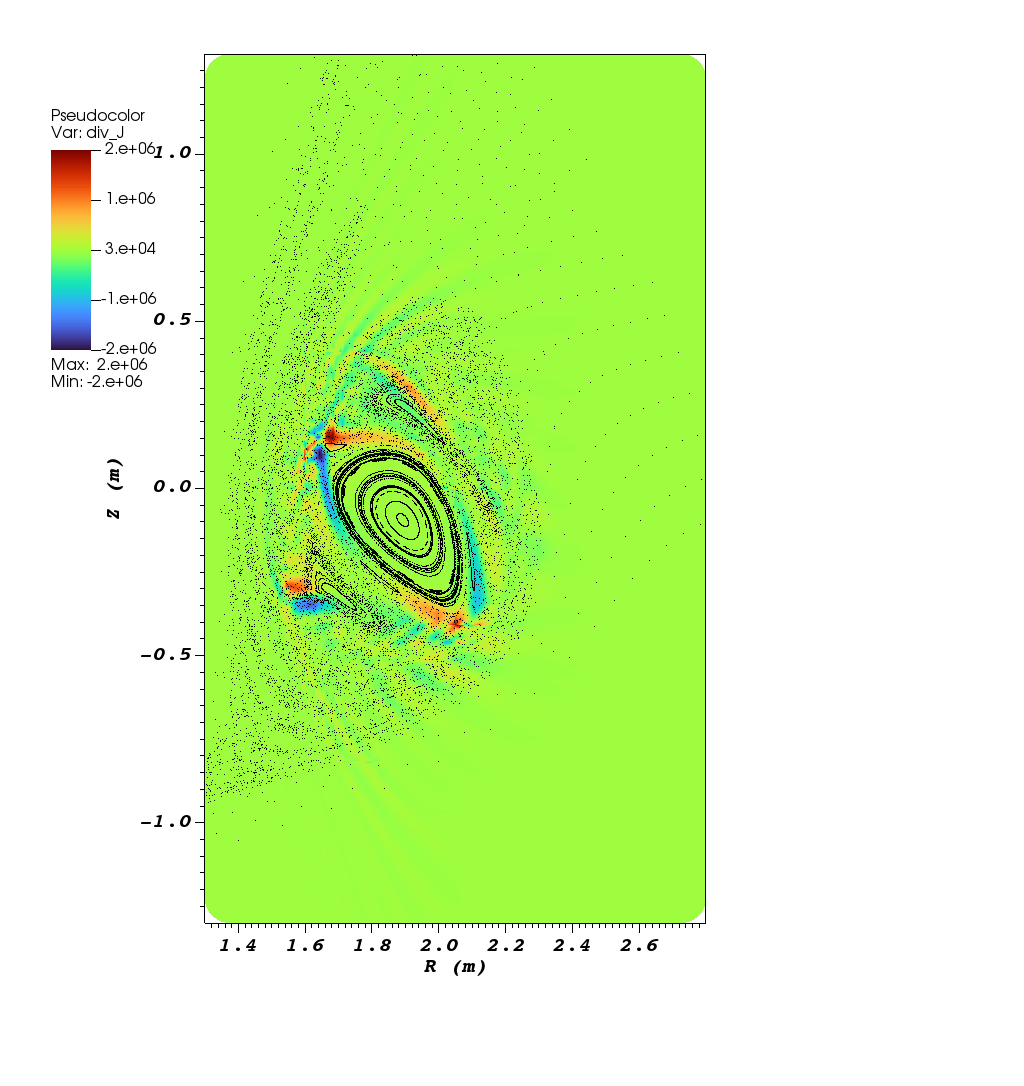}
		\caption{$t=8.8\si{ms}$}
		\label{fig_VDE_rect:div_J_poi_880}
	\end{subfigure}
	\begin{subfigure}[htb]{0.49\textwidth}
		\centering
		\includegraphics[width=1\textwidth]{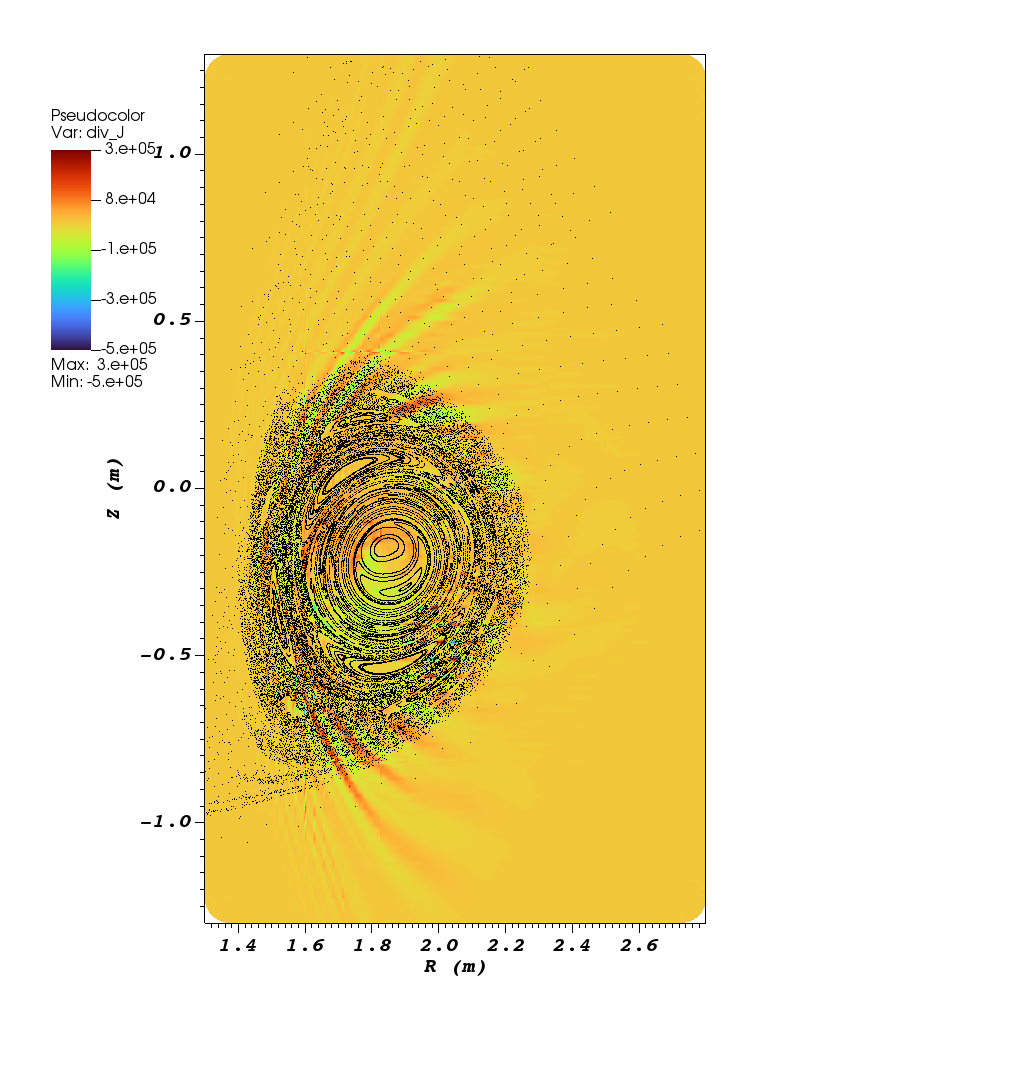}
		\caption{$t=12.2\si{ms}$}
		\label{fig_VDE_rect:div_J_poi_1220}
	\end{subfigure}

\caption{Contours of the divergence of plasma current density and the Poincar{\'e} plots in the poloidal plane at (a) $t=7.2\si{ms}$, (b) $t=8.0\si{ms}$, (c) $t=8.8\si{ms}$, and (d) $t=12.2\si{ms}$ during a non-axisymmetric VDE.}
\label{fig_VDE_rect:div_J_poi}

\end{figure}
\begin{figure}[!htb]
  \centering

	\begin{subfigure}[htb]{0.33\textwidth}
		\centering
		\includegraphics[width=1\textwidth]{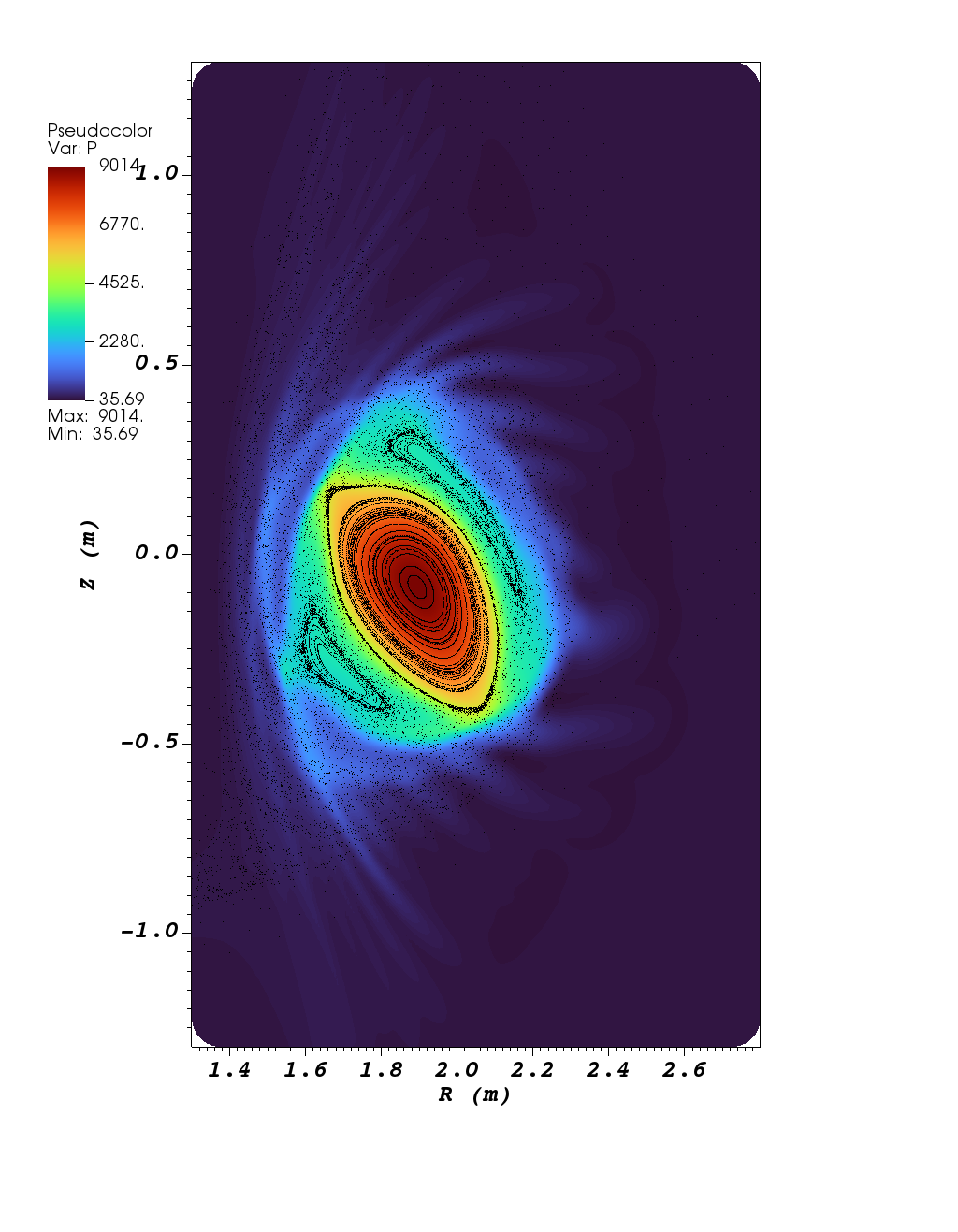}
		\caption{$t=8.40\si{ms}$}
		\label{fig_VDE_rect:quad_poi_840}
	\end{subfigure}\begin{subfigure}[htb]{0.33\textwidth}
		\centering
		\includegraphics[width=1\textwidth]{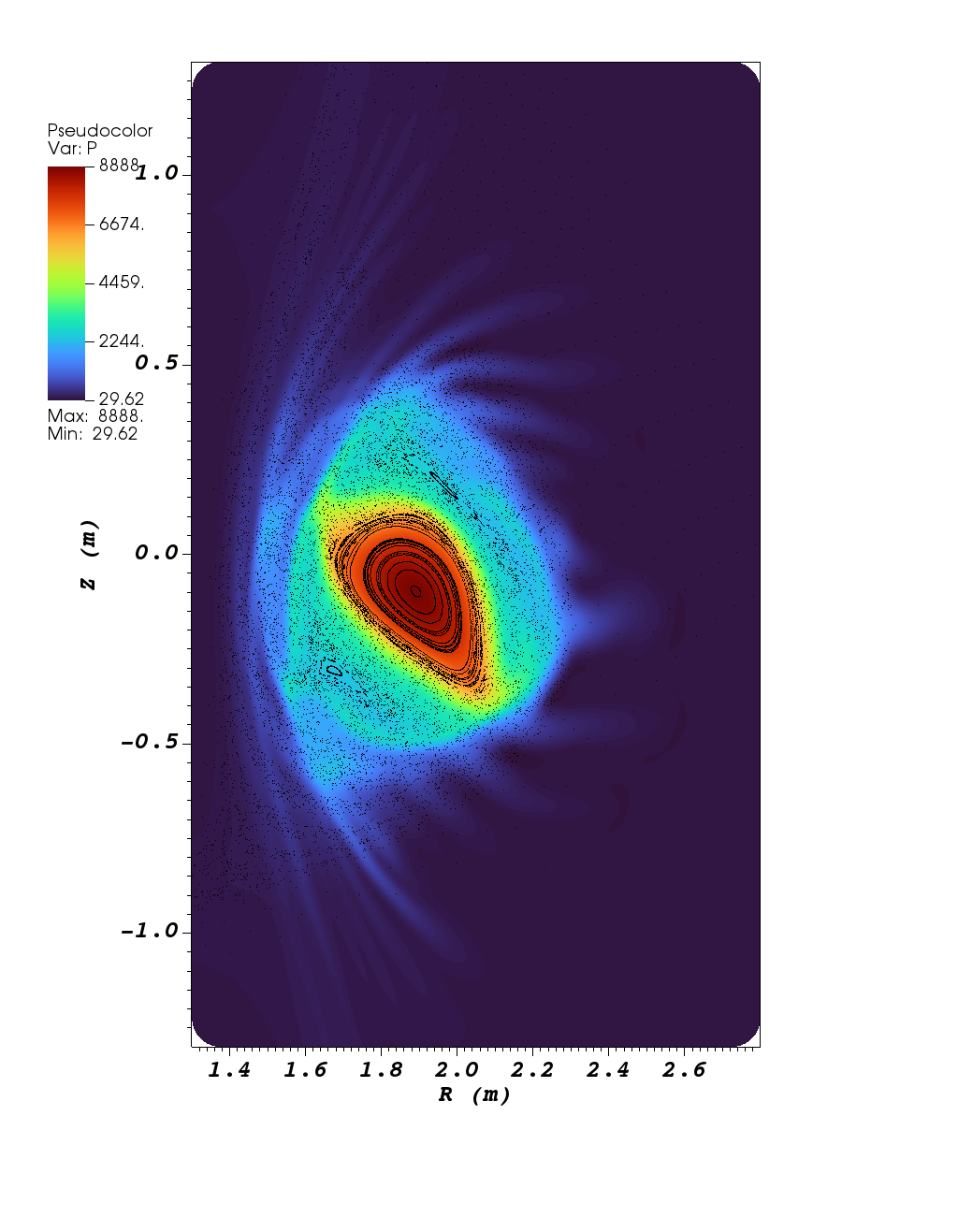}
		\caption{$t=8.94\si{ms}$}
		\label{fig_VDE_rect:quad_poi_894}
	\end{subfigure}
	
	\begin{subfigure}[htb]{0.33\textwidth}
		\centering
		\includegraphics[width=1\textwidth]{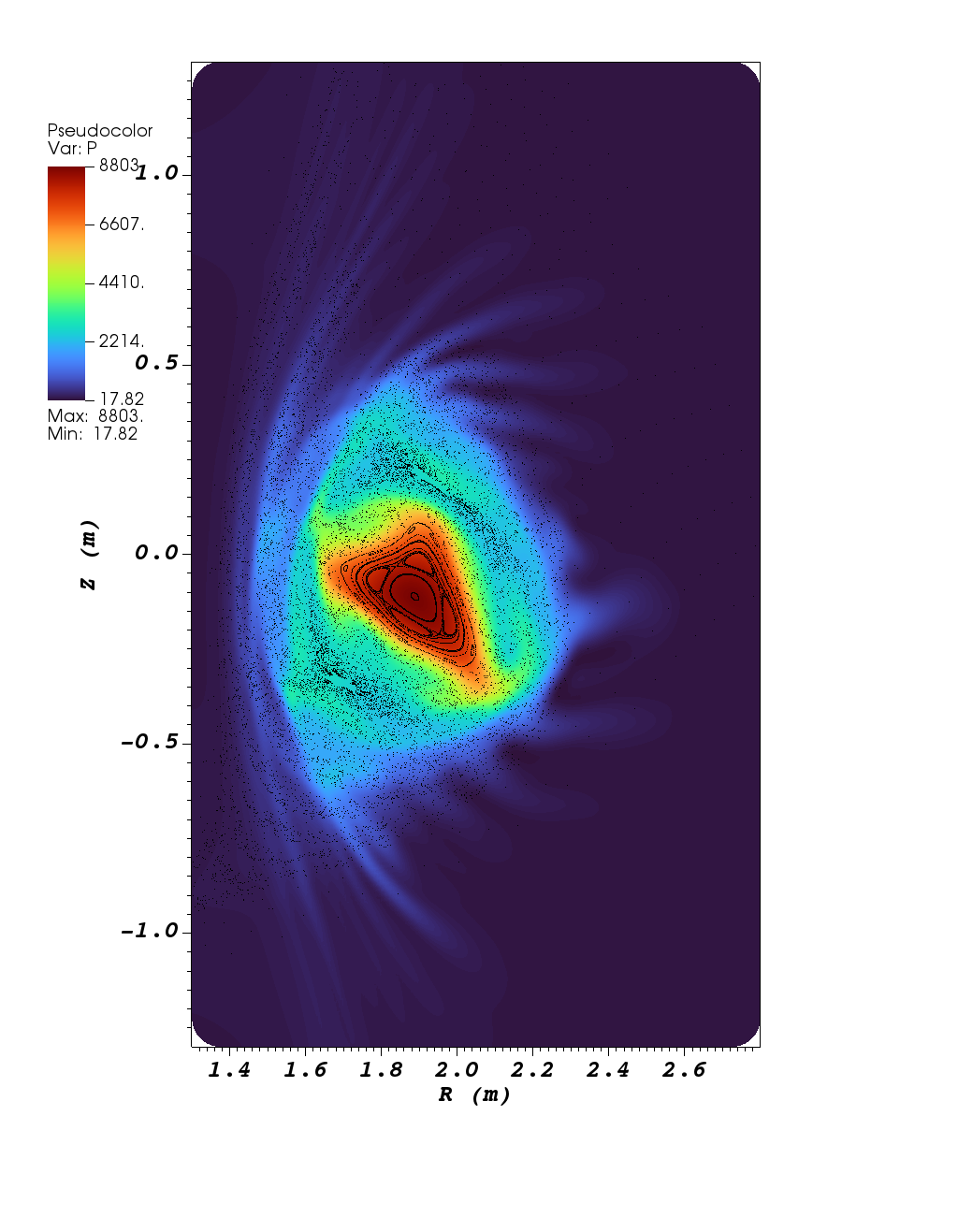}
		\caption{$t=9.30\si{ms}$}
		\label{fig_VDE_rect:quad_poi_930}
	\end{subfigure}\begin{subfigure}[htb]{0.33\textwidth}
		\centering
		\includegraphics[width=1\textwidth]{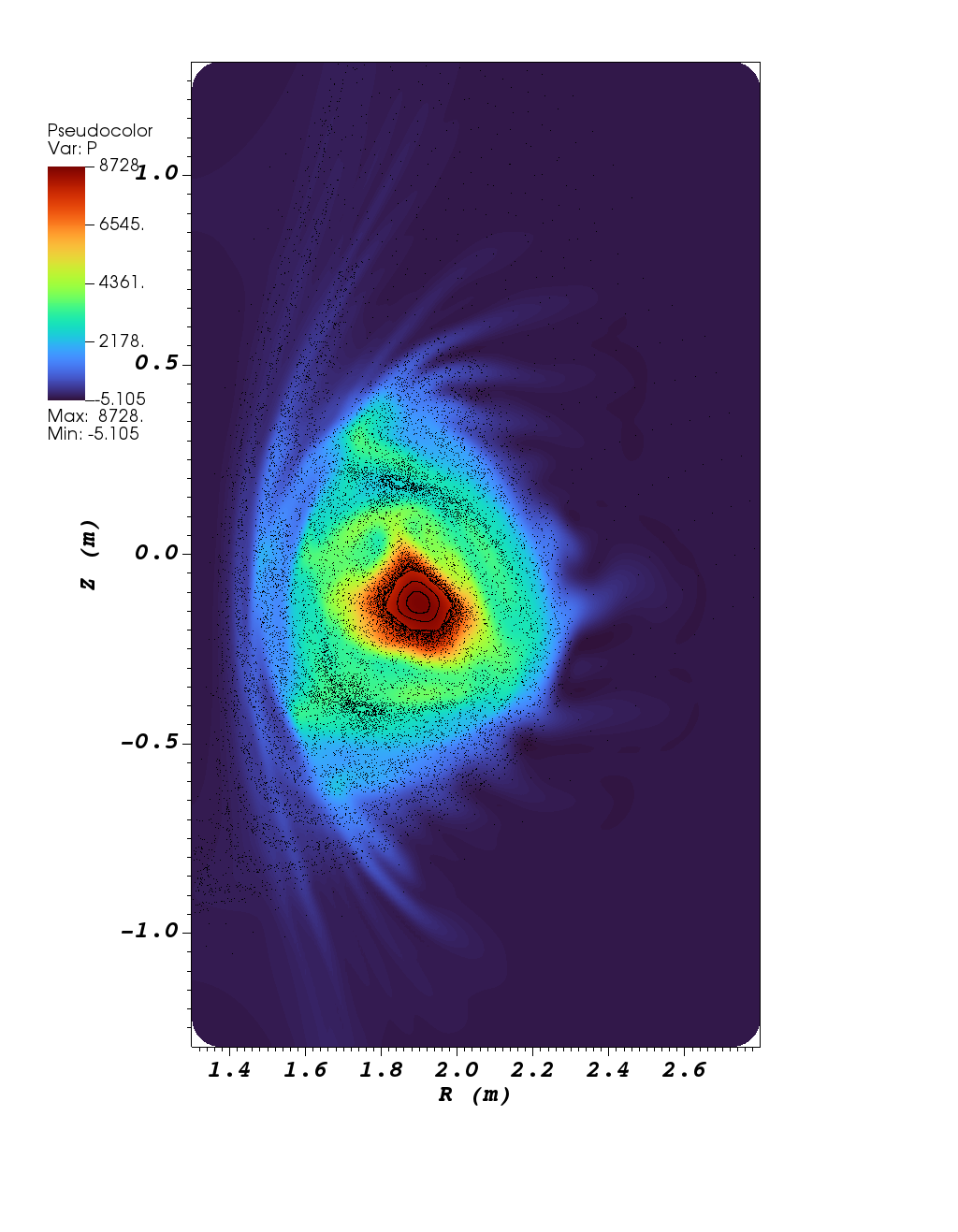}
		\caption{$t=10.00\si{ms}$}
		\label{fig_VDE_rect:quad_poi_1000}
	\end{subfigure}
	
	\begin{subfigure}[htb]{0.33\textwidth}
		\centering
		\includegraphics[width=1\textwidth]{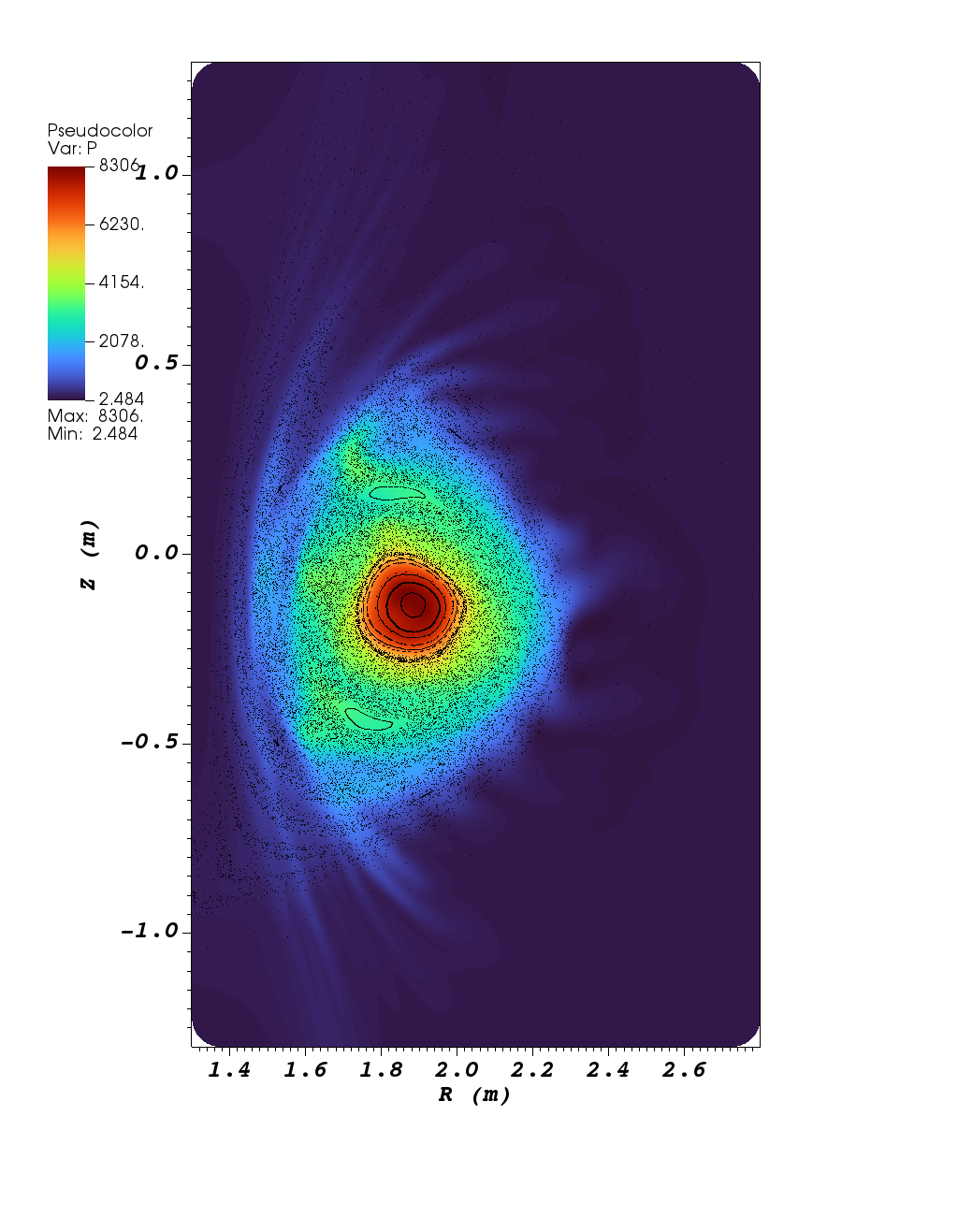}
		\caption{$t=10.60\si{ms}$}
		\label{fig_VDE_rect:quad_poi_1060}
	\end{subfigure}\begin{subfigure}[htb]{0.33\textwidth}
		\centering
		\includegraphics[width=1\textwidth]{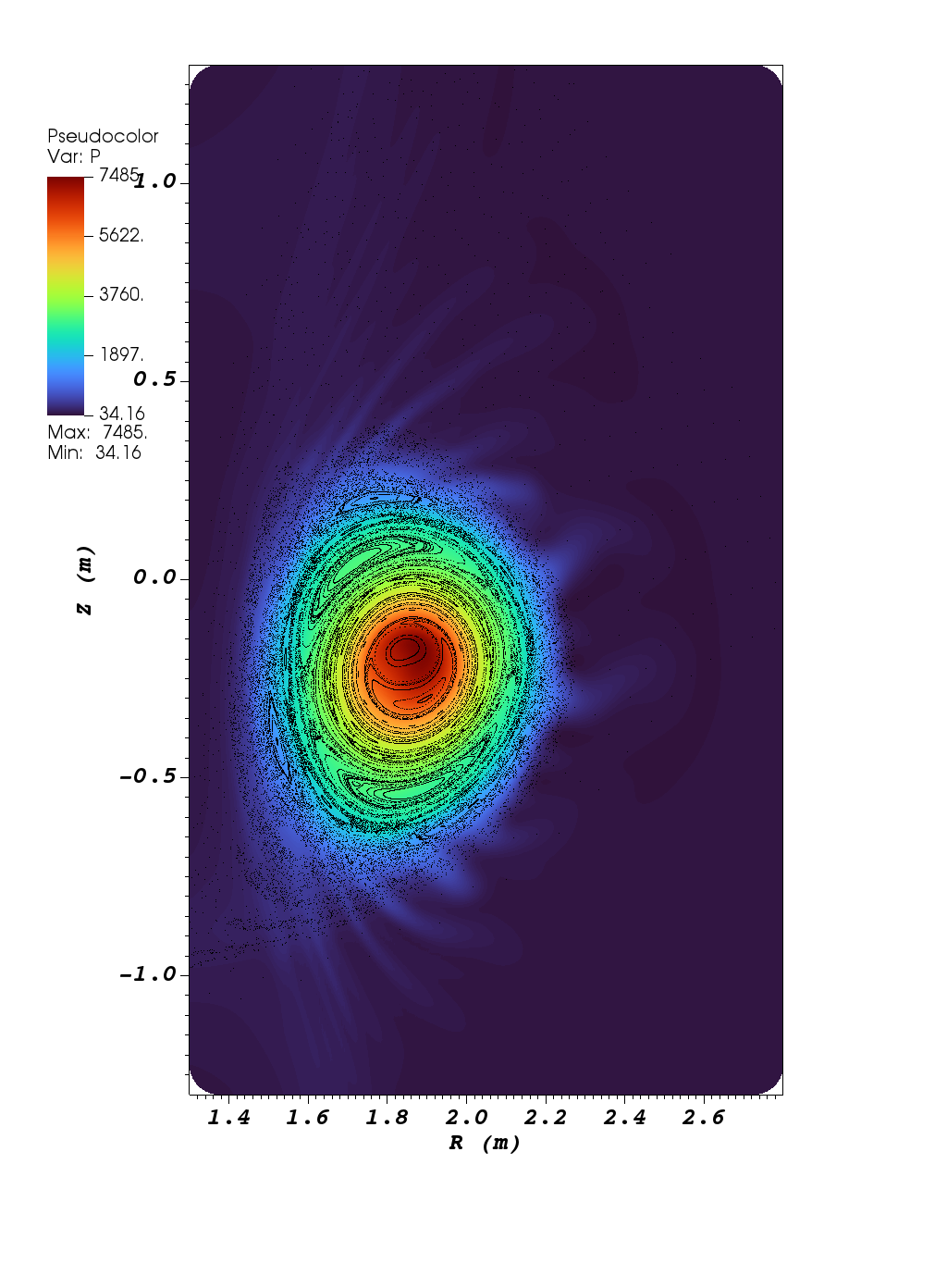}
		\caption{$t=12.26\si{ms}$}
		\label{fig_VDE_rect:quad_poi_1226}
	\end{subfigure}
	
\caption{Contours of the plasma pressure and the Poincar{\'e} plots in the poloidal plane at (a) $t=8.40\si{ms}$, (b) $t=8.94\si{ms}$, (c) $t=9.30\si{ms}$, (d) $t=10.00\si{ms}$, (e) $t=10.60\si{ms}$, and (f) $t=12.26\si{ms}$ during a non-axisymmetric VDE.}
\label{fig_VDE_rect:quad_poi_recovered}

\end{figure}

During the development of an axisymmetric VDE, all magnetic surfaces remain topologically intact without the formation of any magnetic island. In presence of non-axisymmetric toroidal components, the initially formed island chain near edge gradually moves inward to the core region (Fig.\ref{fig_VDE_rect:quad_poincare_250}), until almost all the magnetic flux surfaces are destroyed and the minor disruption starts after the saturation of the non-axisymmetric modes around $t=10 \si{ms}$ (Fig.\ref{fig_VDE_rect:div_J_poi}). There is a strong correlation between the magnetic island location and the divergence of current density (Fig.\ref{fig_VDE_rect:div_J_poi}). After the $(2,1)$ magnetic island formation, the magnetic flux surface becomes healed by $t= 10.8 \si{ms}$ (Fig.\ref{fig_VDE_rect:quad_poi_recovered}). There is a poloidal mode transition from $m=2$ to $m=3$ during the destruction and recovery of the magnetic flux surfaces in the core region between $t=8.4 \si{ms}$ and $t=9.3 \si{ms}$ (Fig.\ref{fig_VDE_rect:quad_poi_recovered}). After the healing of the magnetic flux surface, the $(2,1)$ magnetic island persists along with the VDE process until the plasma touches the wall. Meanwhile, the finger filement structures correlated with the divergence of current density are present throughout the VDE process (Fig.\ref{fig_VDE_rect:div_J_poi}).

\begin{figure}[htb]
  \centering
  \includegraphics[width=1\textwidth]{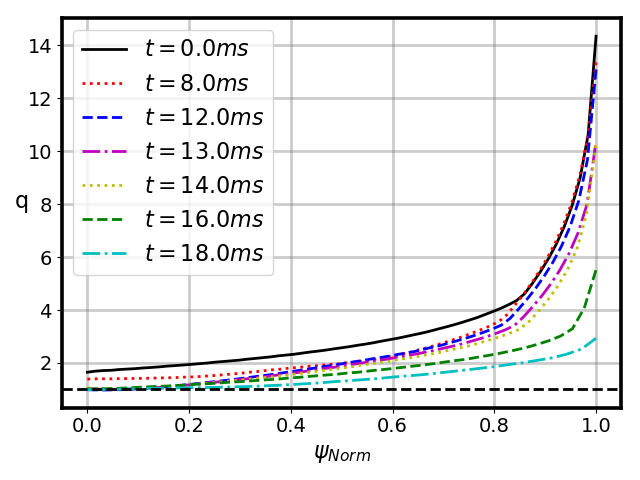}
\caption{Evolution of the safety factor profile during an non-axisymmetric VDE process. The horizontal dashed line stands for $q=1$. The $\psi_{Norm}$ is the normalized poloidal flux.}
  \label{fig_VDE_rect:q_profs}
\end{figure}
\begin{figure}[htb]
  \centering
  \includegraphics[width=1\textwidth]{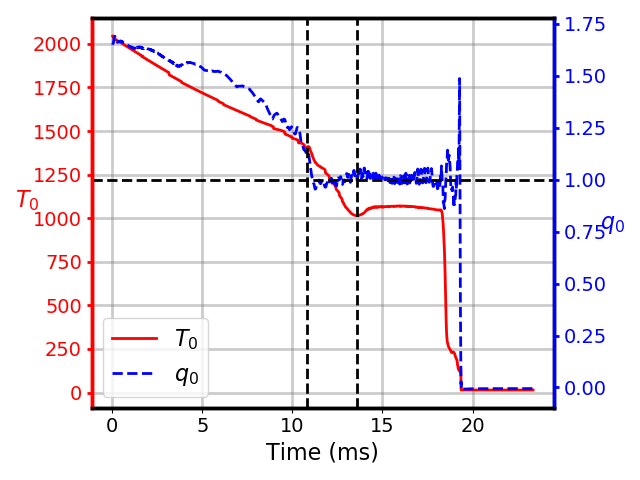}
\caption{The evolution of temperature and safety factor at magnetic axis over time during a non-axisymmetric VDE. The two vertical dashed lines stand for $t_{1}=10.8\si{ms}$ and $t_{2}=13.6\si{ms}$ respectively.}
  \label{fig_VDE_rect:compare_T0_q0}
\end{figure}
\begin{figure}[htb]
  \centering
  \includegraphics[width=1\textwidth]{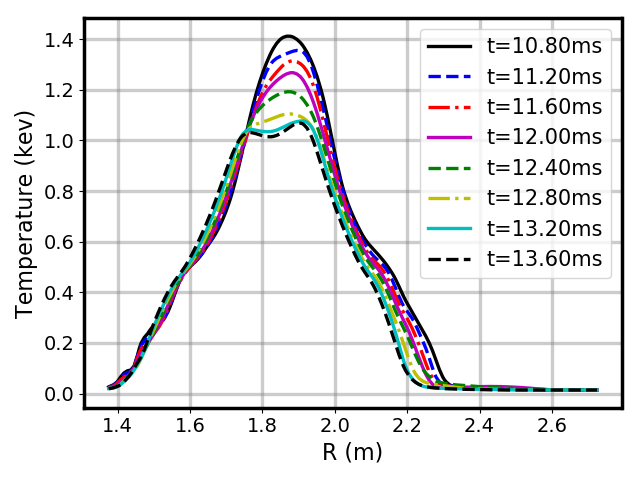}
\caption{The evolution of temperature profile along the line $Z=Z_{\rm{ax}}$ during the thermal quench phase. $Z_{\rm{ax}}$ stands for the vertical coordinate of magnetic axis here.}
  \label{fig_VDE_rect:Ti_profs_kink}
\end{figure}

The edge safety factor drops along with the non-axisymmetric VDE process, which is consistent with the analytic theory (Fig.\ref{fig_VDE_rect:q_profs})~\cite{artola_understanding_2020}. Meanwhile, the the safety factor at magnetic axis $q_{0}$ drops as well during both the axisymmetric and the non-axisymmetric VDE processes (Figs.\ref{fig_VDE_rect:3D_Eth_I_q0} and \ref{fig_VDE_rect:q_profs}). Whereas in the axisymmetric VDE $q_{0}$ remains above unity along with the whole process before the $q=1$ resonant surface touches the wall at $t= 30.5\si{ms}$ , $q_{0}$ in the non-axisymmetric VDE is drops below and oscillates around unity after the minor disruption completes and the magnetic flux surface heals. Thus, the core plasma behaves like a $(1,1)$ internal kink mode because of the oscillation of $q_{0}$ around unity, which leads to the collapse of temperature at the magnetic axis ($T_{0}$) from $t_{1}=10.8 \si{ms}$ to $t_{2}=13.6 \si{ms}$ (Fig.\ref{fig_VDE_rect:compare_T0_q0}. As a result, the core temperature profile flattens  inside the $q=1$ surface during the same period (Figs.\ref{fig_VDE_rect:halo_TPF} and \ref{fig_VDE_rect:Ti_profs_kink}). A $(1,1)$ magnetic island forms in the core region and replaces the magnetic axis during the oscillation of $q_{0}$ around unity over the time interval $11 \si{ms} \leq t \leq 18 \si{ms}$ before the plasma touches the wall (Fig.\ref{fig_VDE_rect:3D_Eth_I_q0}). 

The halo current density $J_{h}$, halo current $I_{h}$, and the toroidal peak factor ($\rm{TPF}$) are defined as: 
\begin{eqnarray}
\label{eqn:J_h}
J_{h}=\vec{J}_{p} \cdot \vec{n} \\
\label{eqn:I_h}
I_{h} (\phi) = \oint R |J_{h}| \rm{d} \emph{l} \\
\label{eqn:TPF}
\rm{TPF} = \frac{\emph{I}_{h}(\phi)_{\rm{max}}}{<\emph{I}_{h}>}
\end{eqnarray}
where $J_{p}$ denotes the poloidal component of current density, $\vec{n}$ denotes the unit vector normal to the magnetic surface $\vec{n}=\frac{\vec{B}_{p} \times \vec{J}_{p}}{|\vec{B}_{p} \times \vec{J}_{p}|}$, $R$ denotes the major radius, $l$ denotes the poloidal arc length along the magnetic flux surface, the integral in  Eq.(\ref{eqn:I_h}) perfoms around the whole poloidal cross-section at any given toroidal angle $\phi$, TPF represents the level of toroidal asymmetry in the halo current, and $<I_{h}>$ denotes the toroidal average of $I_{h}$.

\begin{figure}[htb]
  \centering
  \includegraphics[width=1\textwidth]{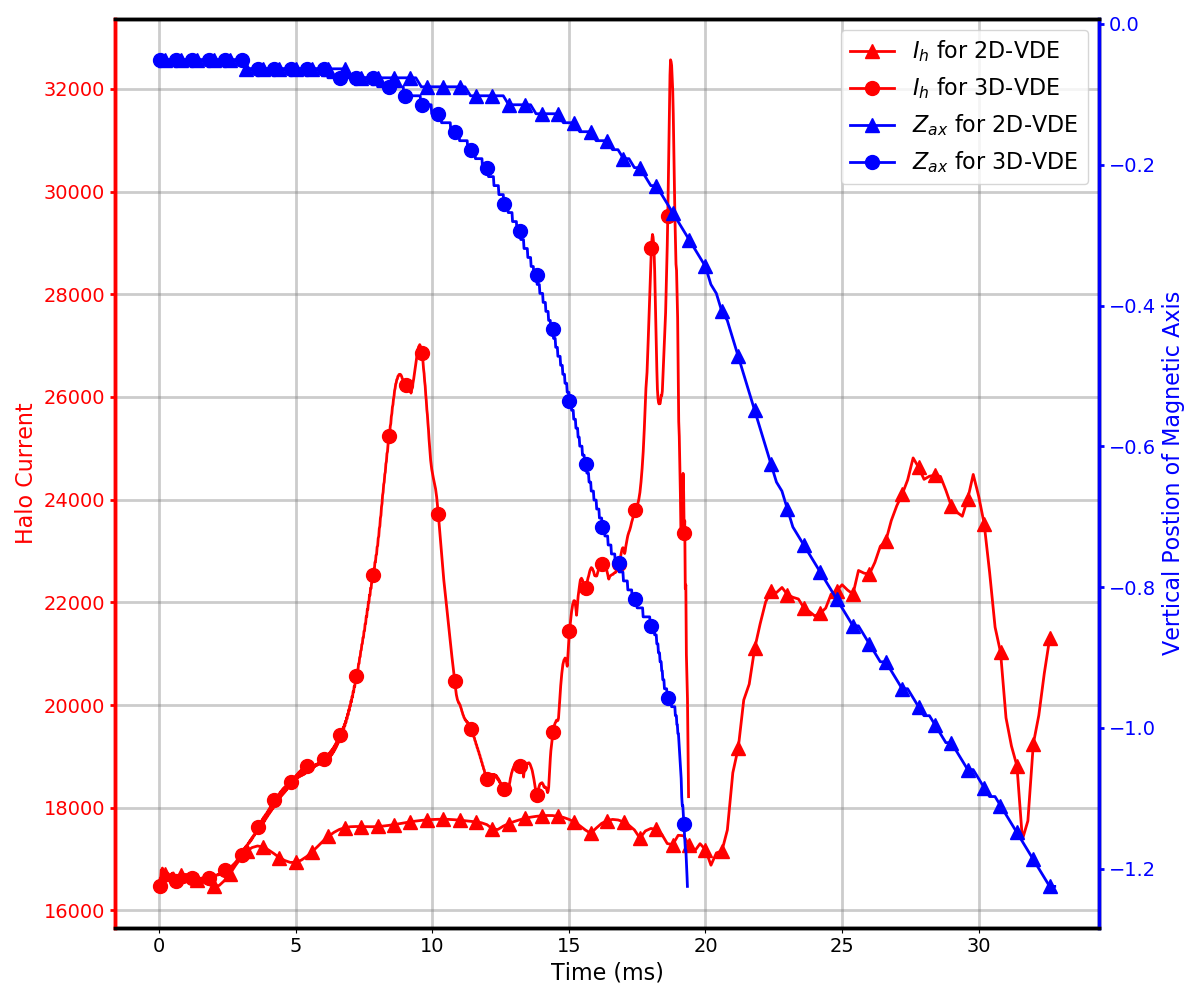}
\caption{The time evolution of halo current (red) and vertical displacement of magnetic axis (blue) during an axisymmetric VDE (2D-VDE, $\blacktriangle$) and a non-axisymmetric VDE (3D-VDE, $\bullet$).}
  \label{fig_VDE_rect:compare_Ih_Z_2D_3D}
\end{figure}
\begin{figure}[!htb]
  \centering

	\begin{subfigure}[htb]{0.49\textwidth}
		\centering
		\includegraphics[width=1\textwidth]{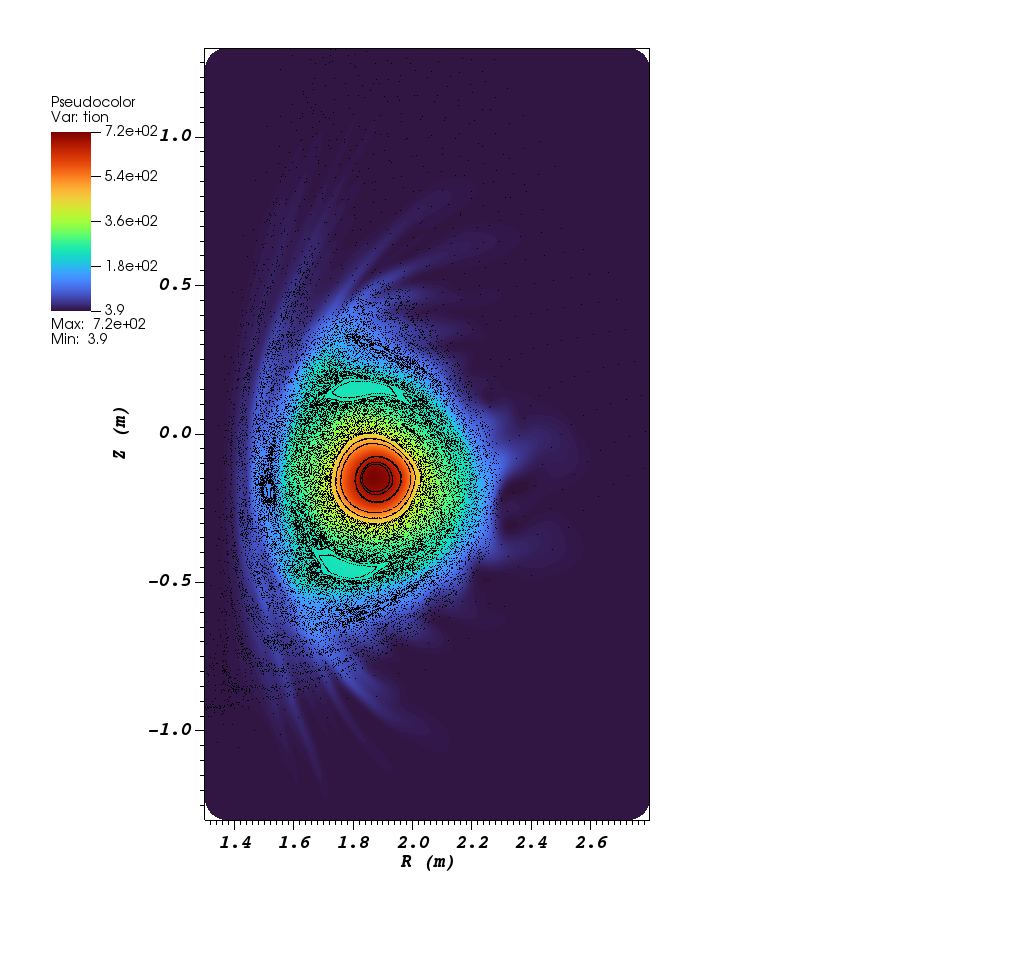}
		\caption{$t=10.8\si{ms}$}
		\label{fig_VDE_rect:Ti_poi_1080}
	\end{subfigure}
    \begin{subfigure}[htb]{0.49\textwidth}
		\centering
		\includegraphics[width=1\textwidth]{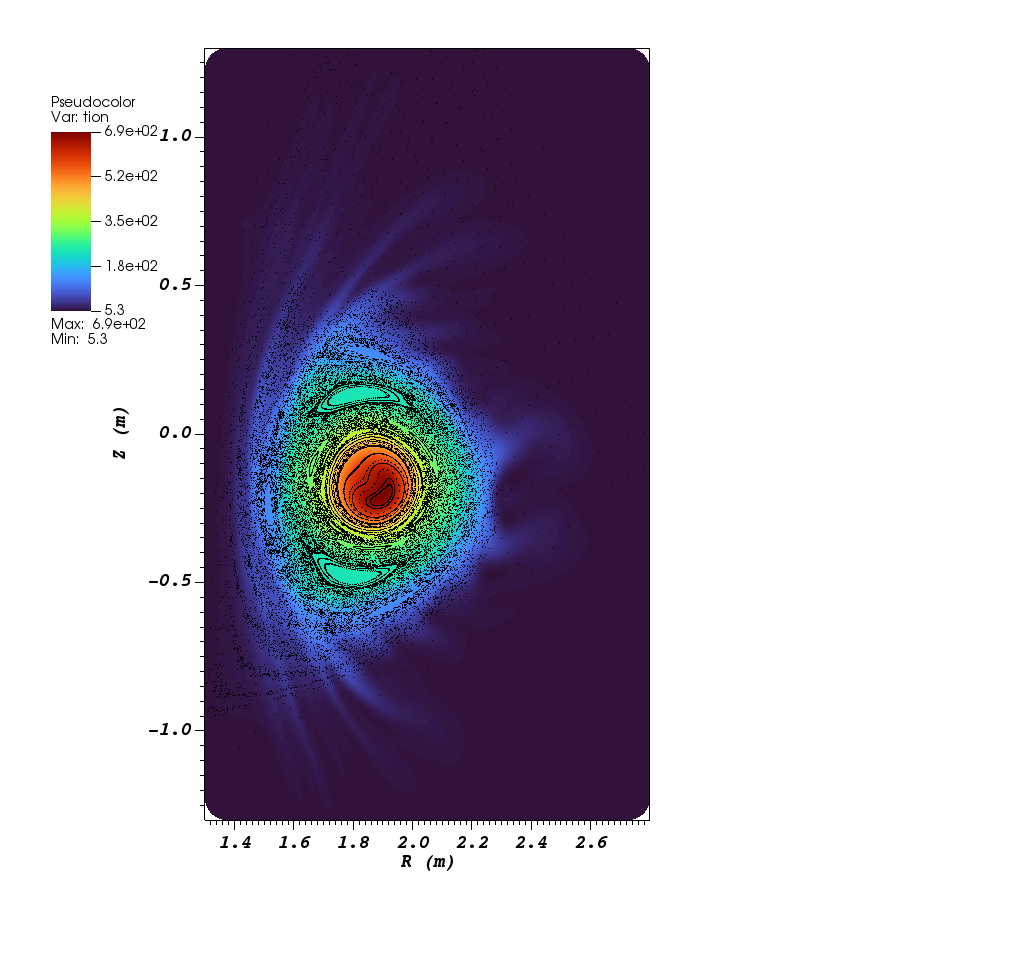}
		\caption{$t=11.2\si{ms}$}
		\label{fig_VDE_rect:Ti_poi_1120}
	\end{subfigure}
	
	\begin{subfigure}[htb]{0.49\textwidth}
		\centering
		\includegraphics[width=1\textwidth]{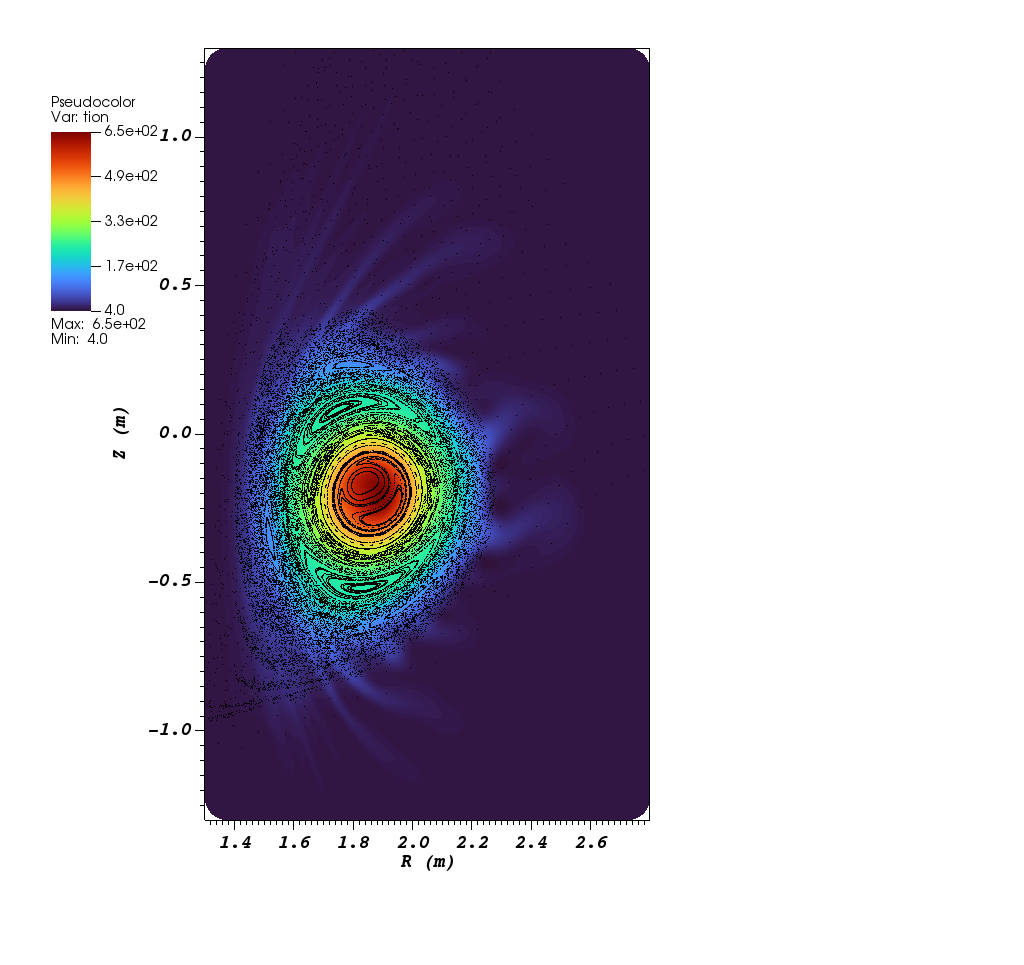}
		\caption{$t=11.9\si{ms}$}
		\label{fig_VDE_rect:Ti_poi_1190}
	\end{subfigure}
	\begin{subfigure}[htb]{0.49\textwidth}
		\centering
		\includegraphics[width=1\textwidth]{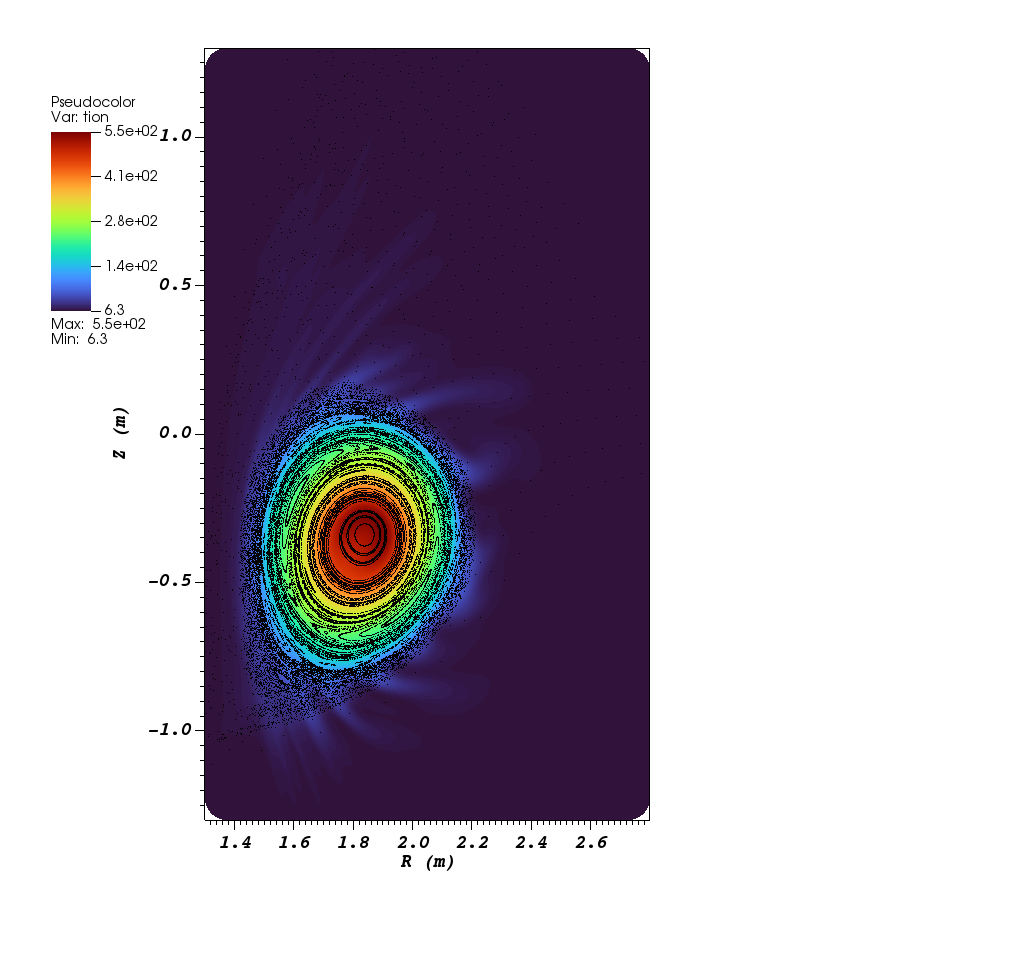}
		\caption{$t=13.8\si{ms}$}
		\label{fig_VDE_rect:Ti_poi_1380}
	\end{subfigure}

\caption{Contours of plasma temperature overlaid with Poincar{\'e} plots in the poloidal plane at (a) $t=10.8 \si{ms}$, (b) $t=11.2 \si{ms}$, (c) $t=11.9 \si{ms}$, and (d) $t=13.8 \si{ms}$ during the thermal quench phase in a non-axisymmetric VDE.}
\label{fig_VDE_rect:Ti_poi_kink}

\end{figure}
\begin{figure}[!htb]
  \centering

	\begin{subfigure}[htb]{0.49\textwidth}
		\centering
		\includegraphics[width=1\textwidth]{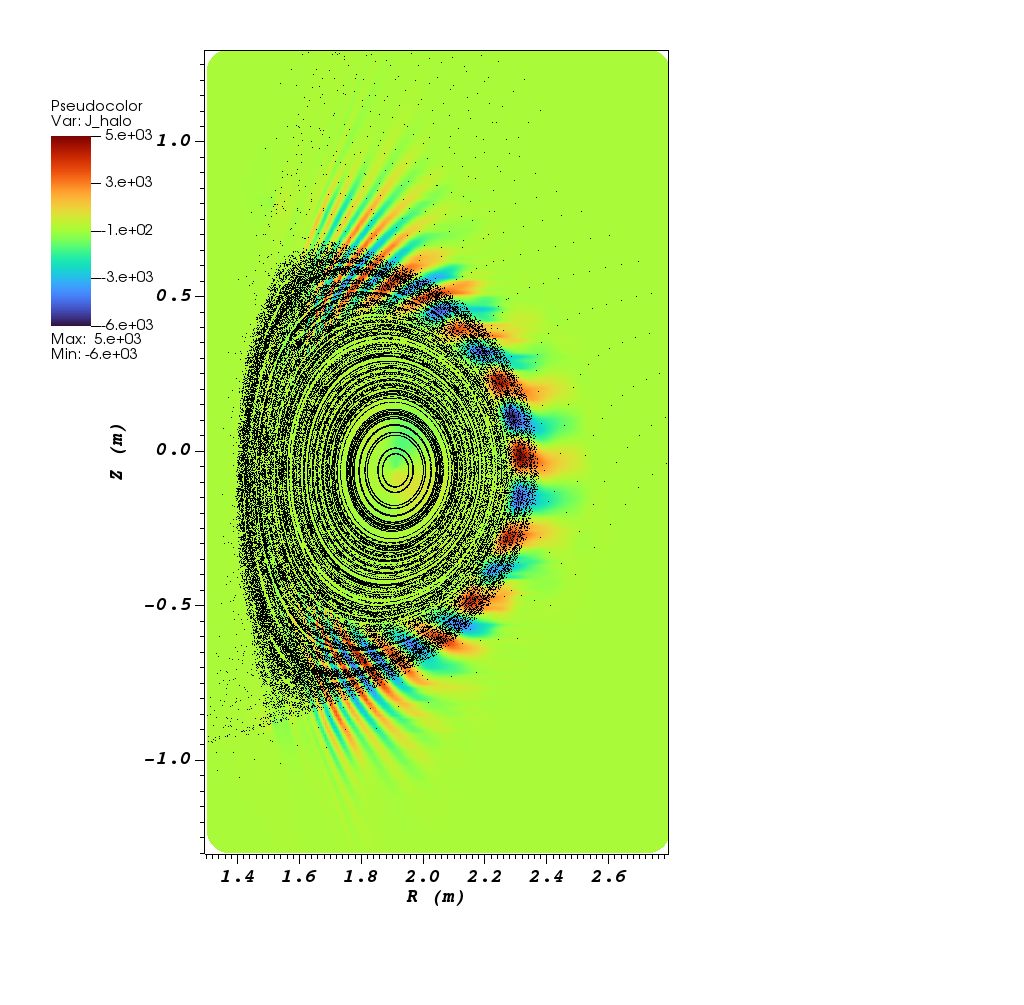}
		\caption{$t=4.4\si{ms}$}
		\label{fig_VDE_rect:halo_poi_440}
	\end{subfigure}
    \begin{subfigure}[htb]{0.49\textwidth}
		\centering
		\includegraphics[width=1\textwidth]{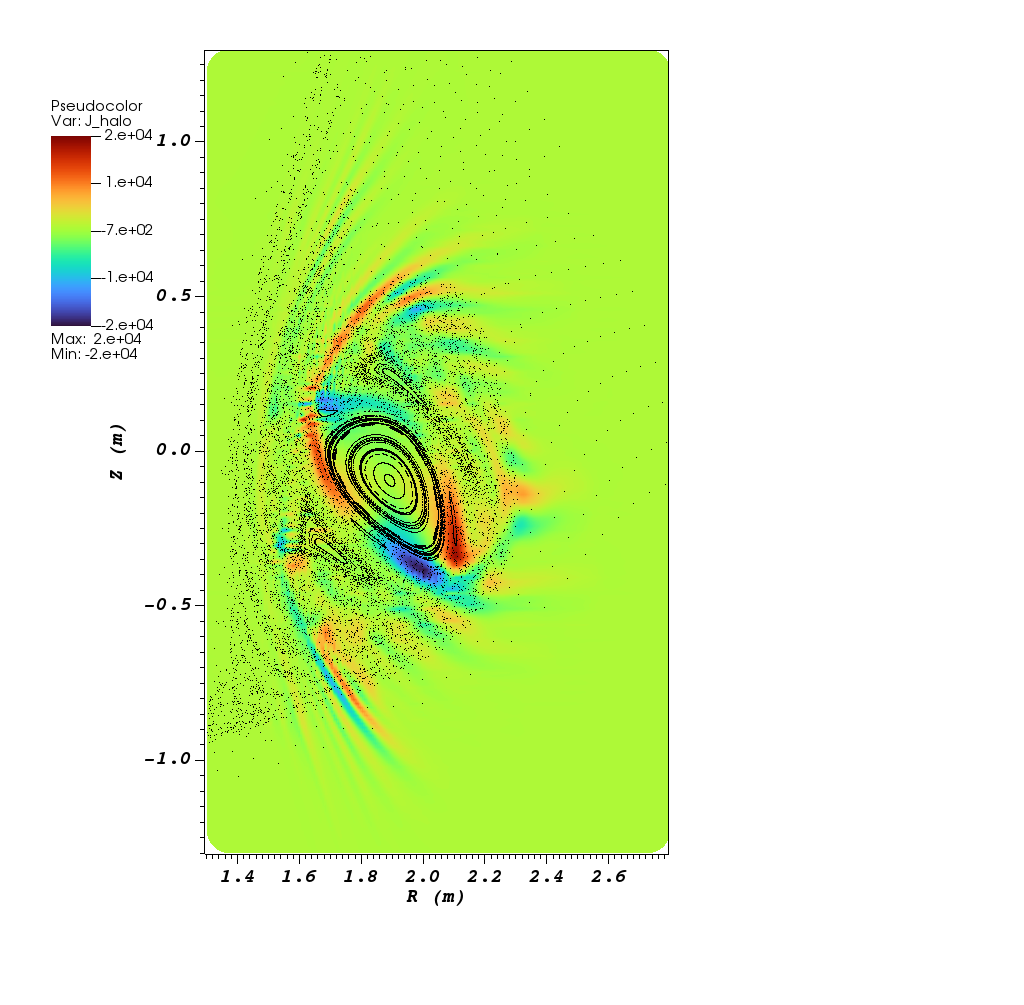}
		\caption{$t=8.8\si{ms}$}
		\label{fig_VDE_rect:halo_poi_880}
	\end{subfigure}
	
	\begin{subfigure}[htb]{0.49\textwidth}
		\centering
		\includegraphics[width=1\textwidth]{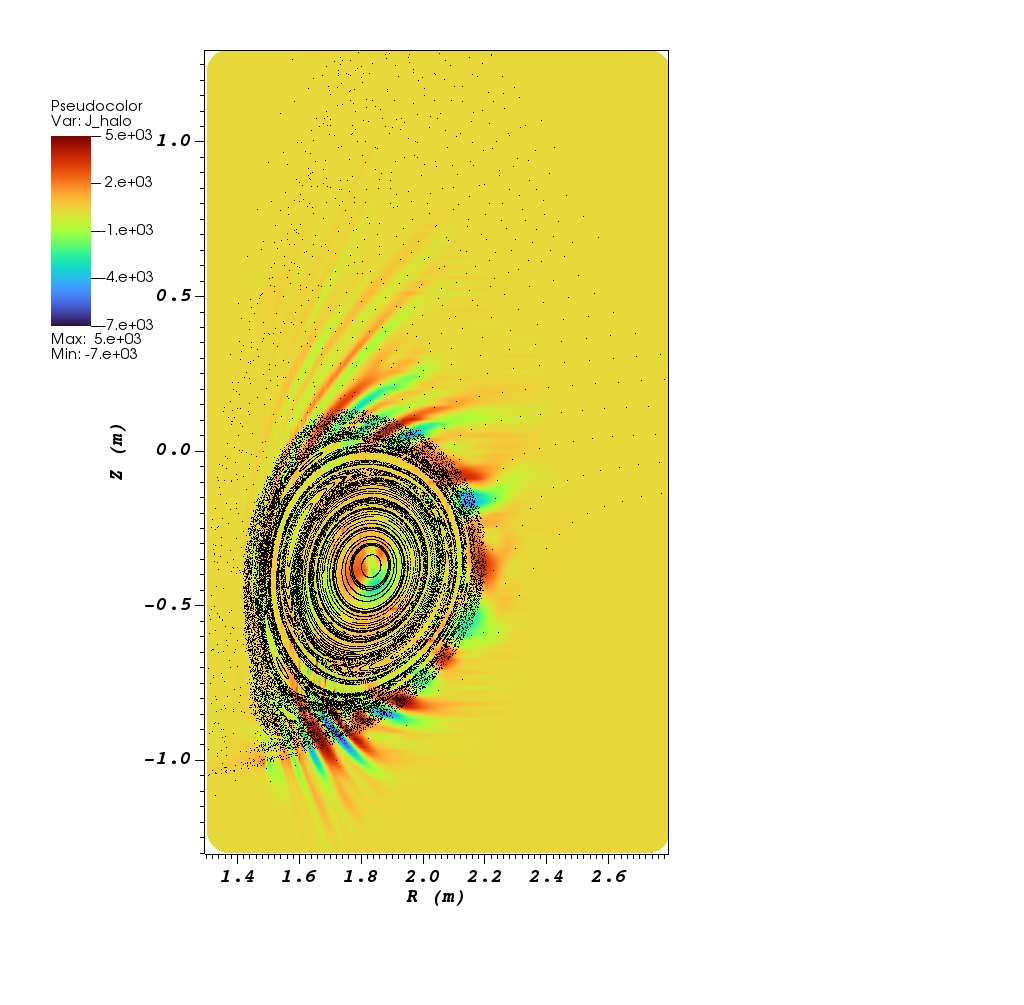}
		\caption{$t=14.0\si{ms}$}
		\label{fig_VDE_rect:halo_poi_1400}
	\end{subfigure}
	\begin{subfigure}[htb]{0.49\textwidth}
		\centering
		\includegraphics[width=1\textwidth]{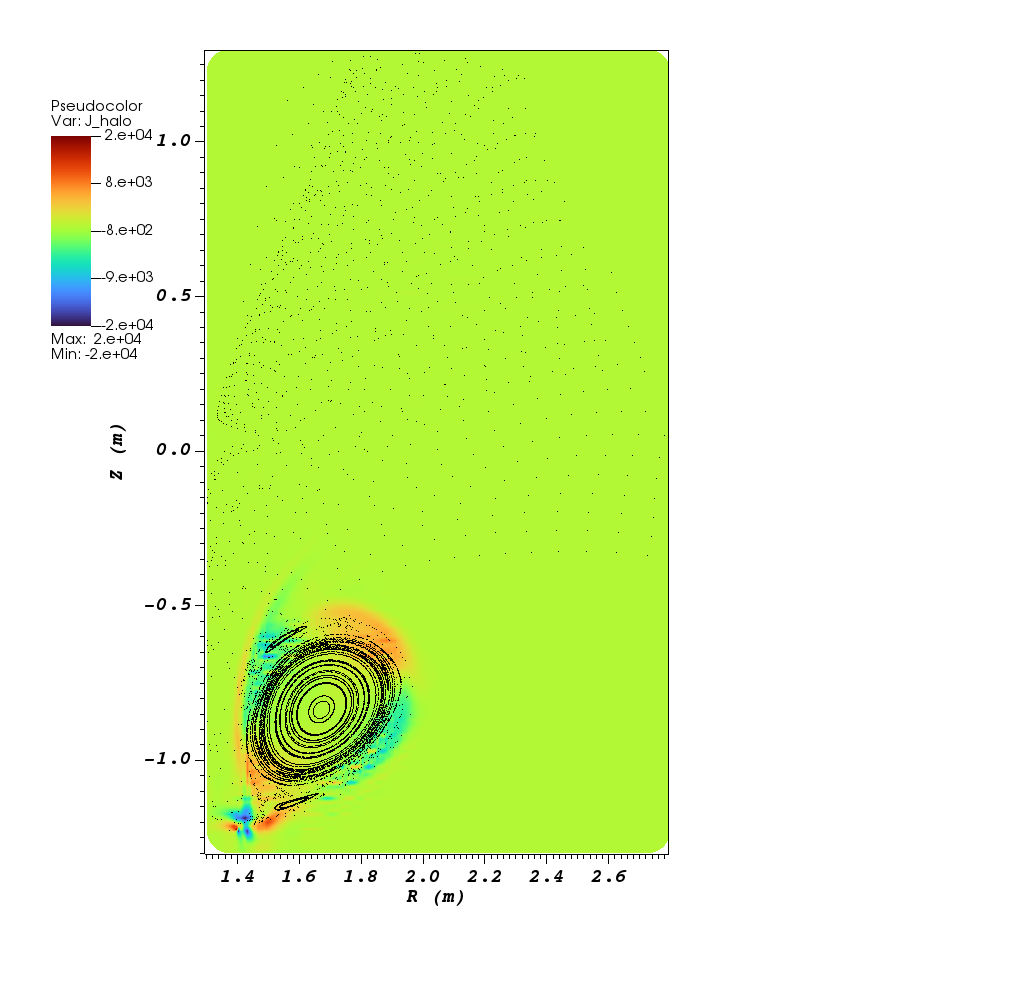}
		\caption{$t=17.6\si{ms}$}
		\label{fig_VDE_rect:halo_poi_1760}
	\end{subfigure}

\caption{Contours of the halo current density and the Poincar{\'e} plots in the poloidal plane at (a) $t=4.4 \si{ms}$, (b) $t=8.8 \si{ms}$, (c) $t=14.0 \si{ms}$, (d) $t=17.6 \si{ms}$ during a non-axisymmetric VDE.}
\label{fig_VDE_rect:halo_poi}

\end{figure}

The halo current is generated when the plasma current flows across the closed flux surfaces during the VDE process and the amplitude of halo current increases rapidly when plasma touches the wall (Fig.\ref{fig_VDE_rect:compare_Ih_Z_2D_3D}). Unlike the axisymmetric VDE, the magnetic island and stochastic field destroy the closed flux surfaces during a non-axisymmetric VDE, which further enhances the poloidal current flows out of the separatrix into the vaccum and the conducting wall when plasma touches the wall. The finger filement structures of halo current density are induced by the formation and destruction of magnetic island chains (Figs.\ref{fig_VDE_rect:halo_poi_440},\ref{fig_VDE_rect:halo_poi_1400}). The destruction of closed flux surface releases the constraint of magnetic tension on the poloidal plasma current, allowing the poloidal plasma current to jet towards the direction of hoop force. Meanwhile, the destruction of closed flux surface also leads to the jet of plasma, which gives rise to the finger and filement structures in plasma pressure in Fig.\ref{fig_VDE_rect:quad_poi_recovered} and temperature in Fig.\ref{fig_VDE_rect:Ti_poi_kink}. Since the formation of halo current depends on the destruction of closed flux surface, the halo current achieves maximum amplitude when the closed flux surfaces are almost all destroyed after the minor as well as the major disruptions (Fig.\ref{fig_VDE_rect:halo_TPF}). An $m=2$ structure in the halo current density is induced by the $(2,1)$ magnetic island during the VDE process after the recovery of magnetic flux surface (Figs.\ref{fig_VDE_rect:halo_poi_880},\ref{fig_VDE_rect:halo_poi_1760}).

\begin{figure}[!htb]
  \centering

	\begin{subfigure}[htb]{0.9\textwidth}
		\centering
		\includegraphics[width=1\textwidth]{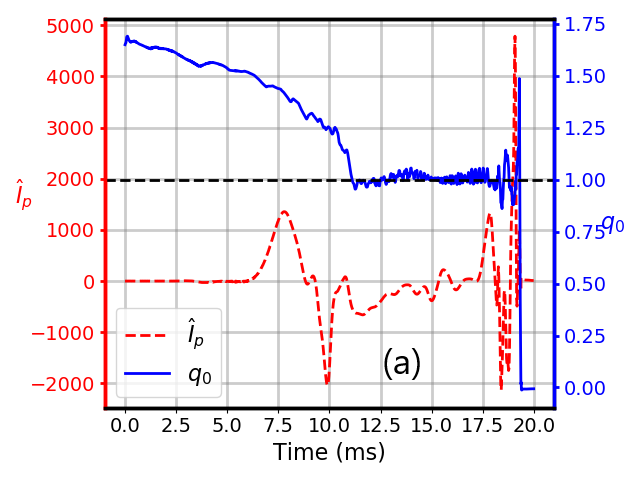}
		\label{fig_VDE_rect:compare_Ih_q0_all}
	\end{subfigure}
	
    \begin{subfigure}[htb]{0.9\textwidth}
		\centering
		\includegraphics[width=1\textwidth]{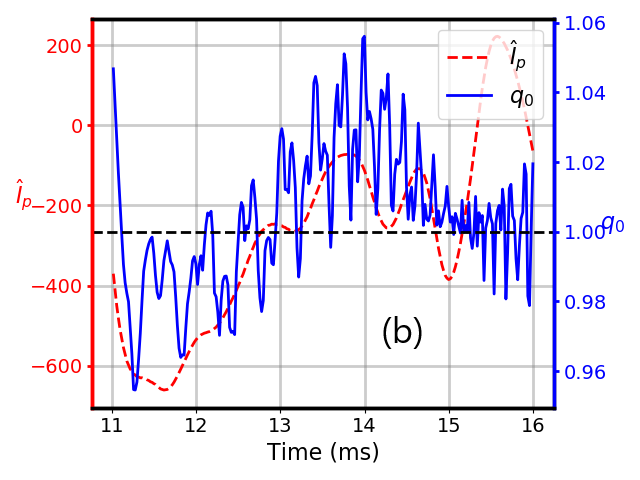}
		\label{fig_VDE_rect:compare_Ih_q0_zoom}
	\end{subfigure}

\caption{(a) Time evolution of the safety factor at magnetic axis $q_{0}$ and the toroidal asymmetry in plasma current $\widehat{I}_{p} (t,0)$ during a non-axisymmetric VDE, and (b) is the zoomed-in part of (a) in the time interval of $11 \si{ms} \leq t \leq 16 \si{ms}$.}
\label{fig_VDE_rect:compare_Ih_q0}

\end{figure}

\begin{figure}[!htb]
  \centering

	\begin{subfigure}[htb]{0.49\textwidth}
		\centering
		\includegraphics[width=1\textwidth]{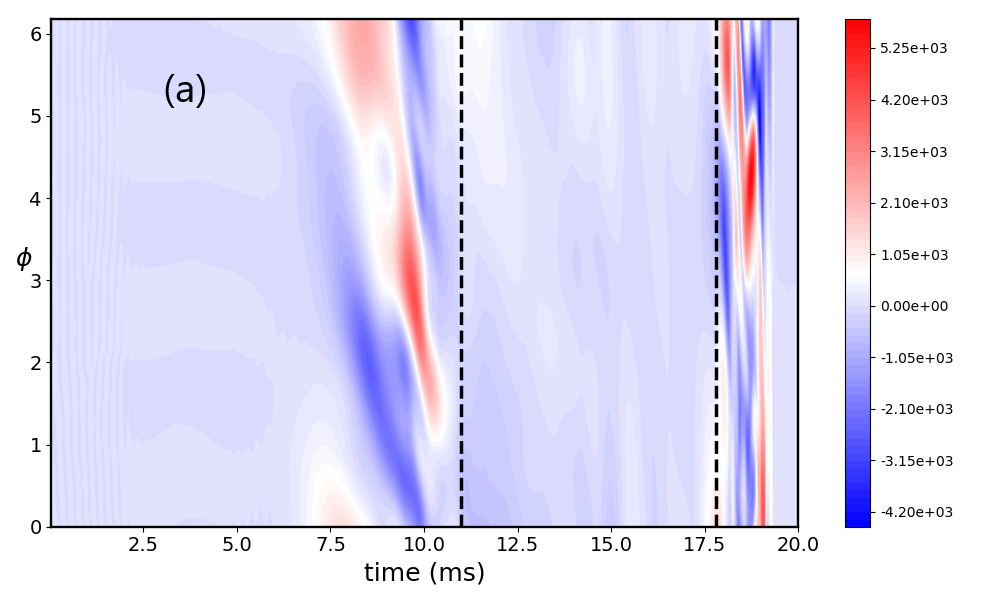}
		\label{fig_VDE_rect:asym_Ip_all}
	\end{subfigure}
    \begin{subfigure}[htb]{0.49\textwidth}
		\centering
		\includegraphics[width=1\textwidth]{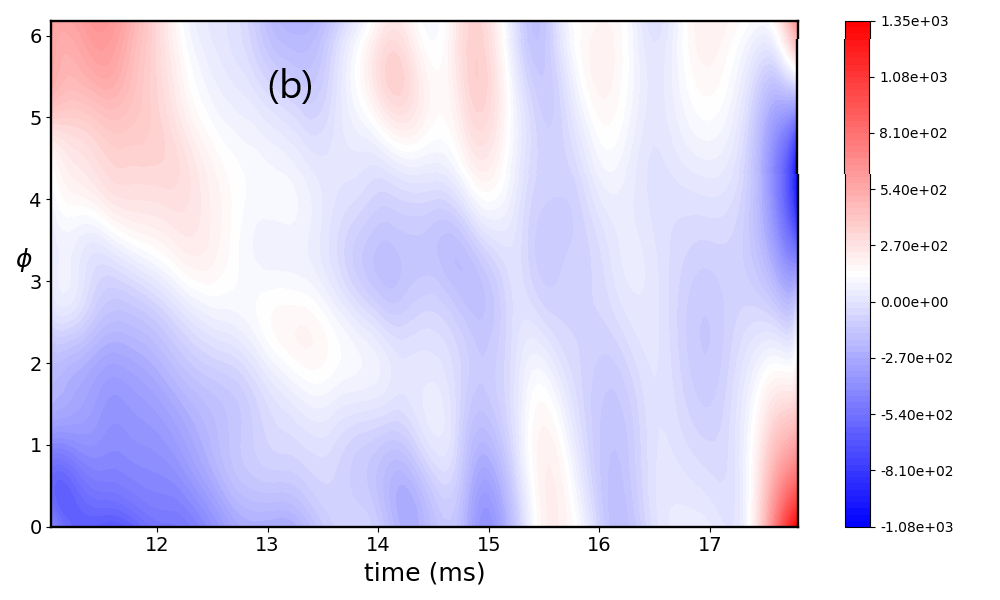}
		\label{fig_VDE_rect:asym_Ip_zoom}
	\end{subfigure}
	
	\begin{subfigure}[htb]{0.49\textwidth}
		\centering
		\includegraphics[width=1\textwidth]{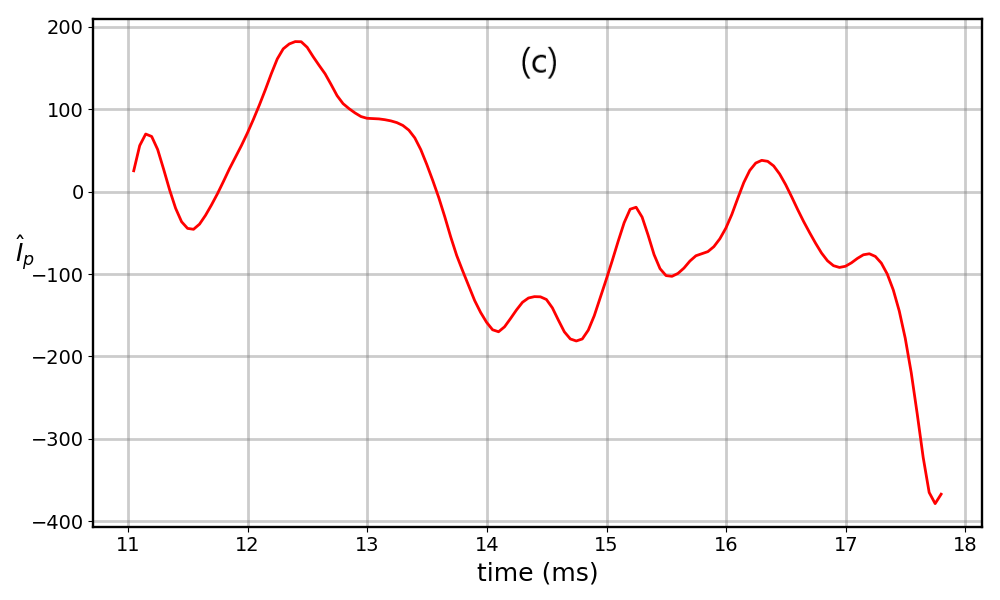}
		\label{fig_VDE_rect:asym_Ip_phi0}
	\end{subfigure}
	\begin{subfigure}[htb]{0.49\textwidth}
		\centering
		\includegraphics[width=1\textwidth]{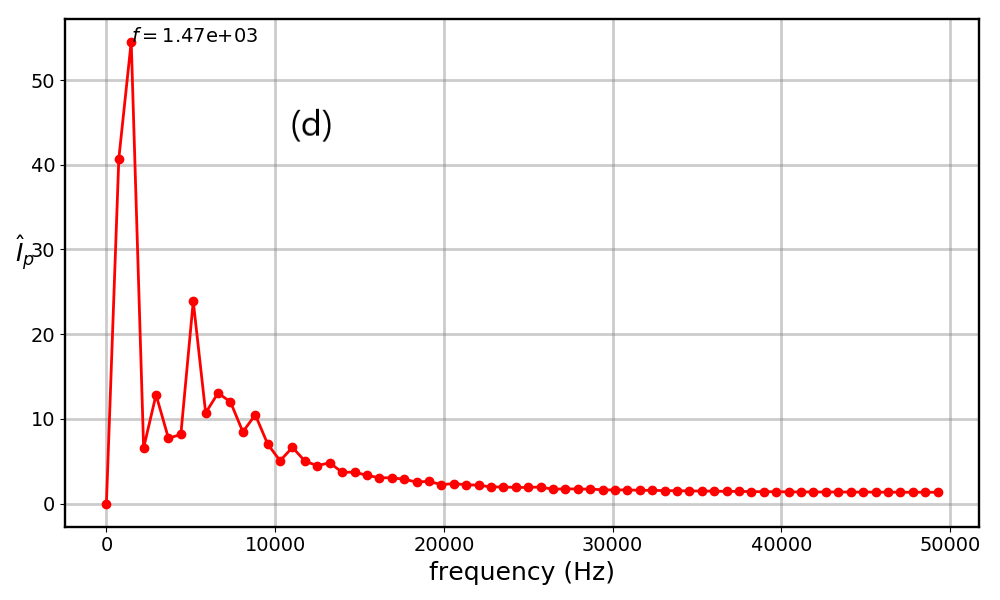}
		\label{fig_VDE_rect:asym_Ip_phi0_fft}
	\end{subfigure}

\caption{(a) Time evolution of the toroidal asymmetry of plasma current $\widehat{I}_{p}(t,\phi)$ at each toroidal angle $\phi$ during a non-axisymmetric VDE, whereas (b) is the zoomed-in part of (a) in the time interval of $11.0 \si{ms} \leq t \leq 17.8 \si{ms}$, (c) is $\widehat{I}_{p} (t,\pi)$, and (d) is the temporal FFT spectrum of $\widehat{I}_{p} (t,\pi)$.}

\label{fig_VDE_rect:asym_Ip}

\end{figure}

The toroidal asymmetry of plasma current $\widehat{I}_{p}(t,\phi)$ is defined as $\widehat{I}_{p} (t,\phi) = I_{p}(t,\phi) - <I_{p}(t,\phi)> $, where $\phi$ denotes the toroidal angle and the  $<I_{p}(t,\phi)> $ denotes the toroidal average in $\phi$~\cite{gerasimov_plasma_2014,saperstein_halo_2022}. The toroidal asymmetry in halo current $\widehat{I}_{h} (\phi)$ is similarly defined as well. The $n=1$ component dominates the toroidal mode structure (Fig.\ref{fig_VDE_rect:asym_Ip}), which is consistent with the perturbation energy history. There is a clear correlation between the oscillation amplitude of $q_{0}$ around unity and that of the asymmetry of plasma current in time (Fig.\ref{fig_VDE_rect:compare_Ih_q0}).

\begin{figure}[!htb]
  \centering
	\begin{subfigure}[htb]{0.49\textwidth}
		\centering
		\includegraphics[width=1\textwidth]{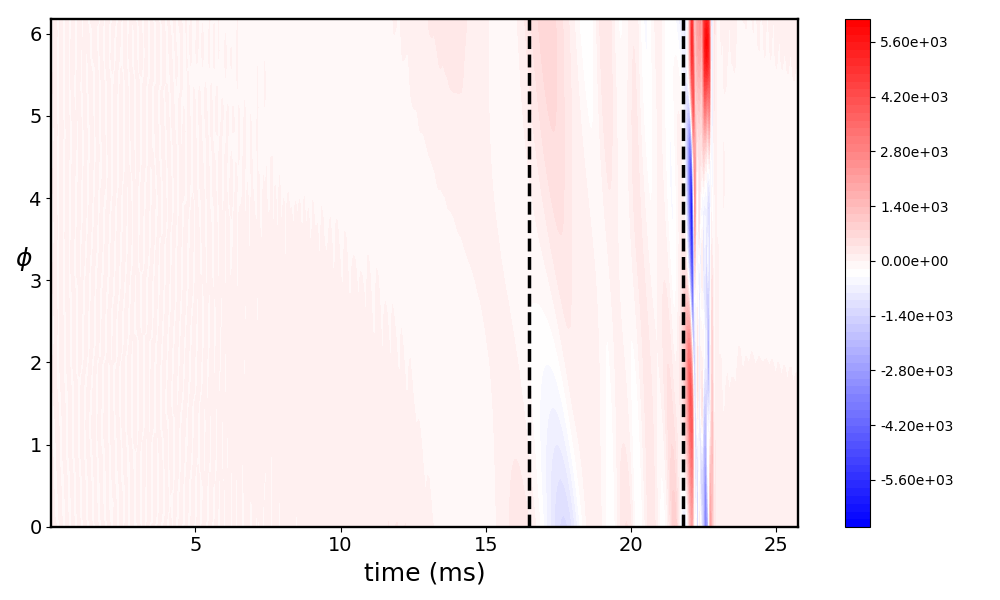}
		\caption{$\chi_{\parallel}/\chi_{\perp}= 10^{7}$}
		\label{fig_VDE_rect:asym_Ip_all_kpll1e7}
	\end{subfigure}
    \begin{subfigure}[htb]{0.49\textwidth}
		\centering
		\includegraphics[width=1\textwidth]{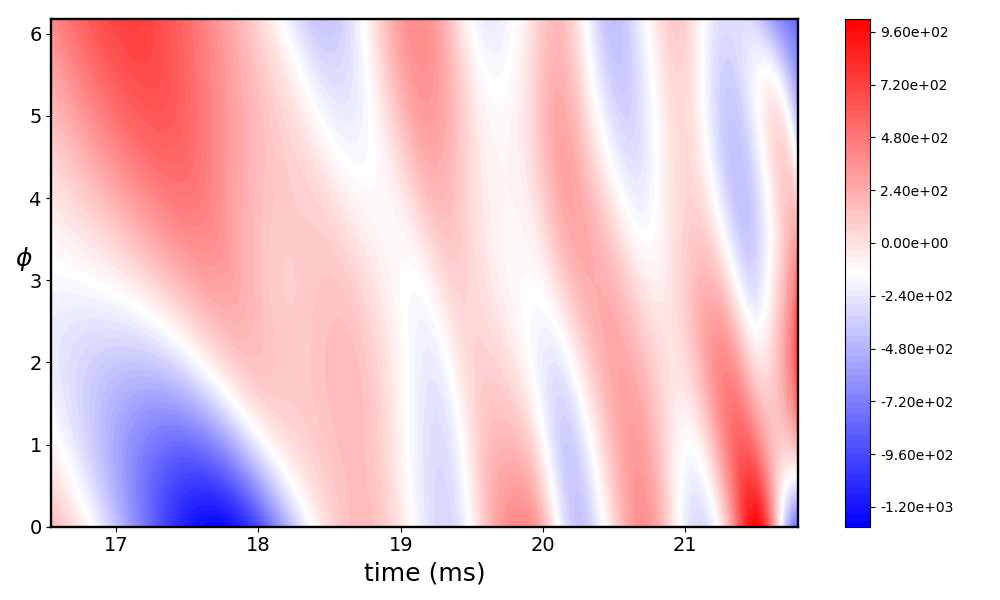}
		\caption{Zoomed-in time interval of (a)}
		\label{fig_VDE_rect:asym_Ip_zoom_kpll1e7}
	\end{subfigure}

	\begin{subfigure}[htb]{0.49\textwidth}
		\centering
		\includegraphics[width=1\textwidth]{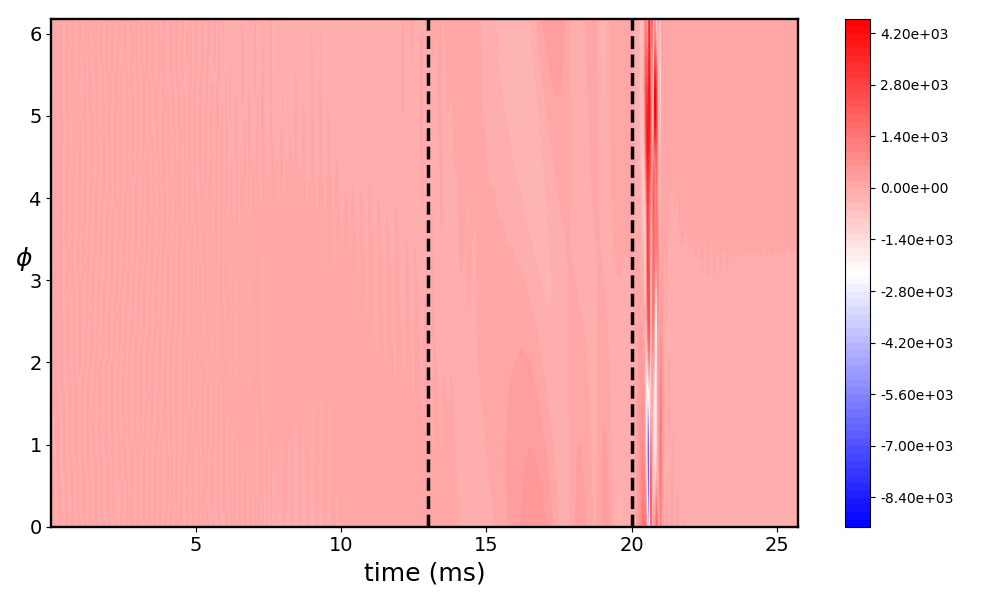}
		\caption{$\chi_{\parallel}/\chi_{\perp}= 10^{8}$}
		\label{fig_VDE_rect:asym_Ip_all_kpll1e8}
	\end{subfigure}
    \begin{subfigure}[htb]{0.49\textwidth}
		\centering
		\includegraphics[width=1\textwidth]{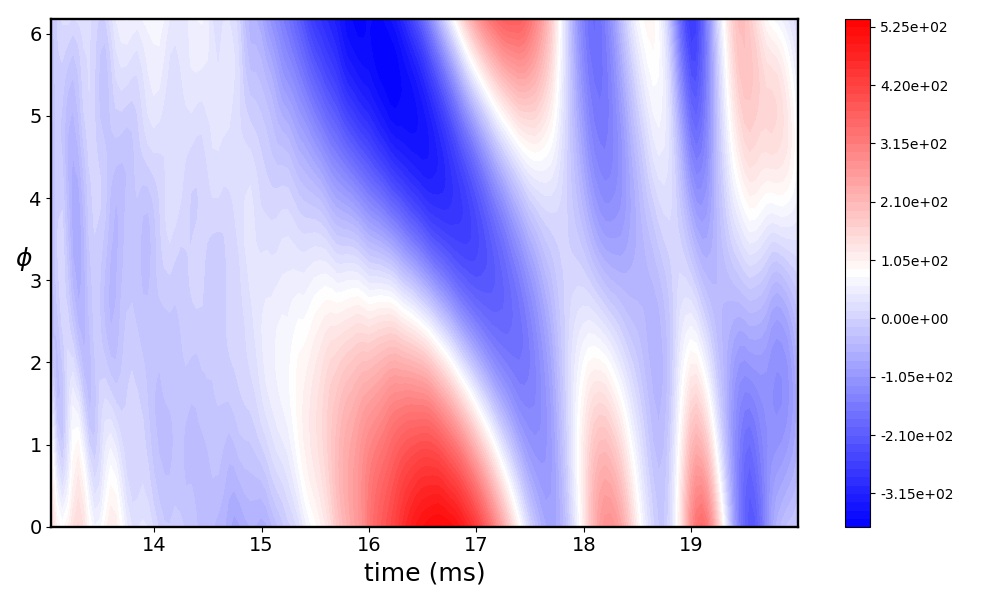}
		\caption{Zoomed-in time interval of (c)}
		\label{fig_VDE_rect:asym_Ip_zoom_kpll1e8}
	\end{subfigure}	

	\begin{subfigure}[htb]{0.49\textwidth}
		\centering
		\includegraphics[width=1\textwidth]{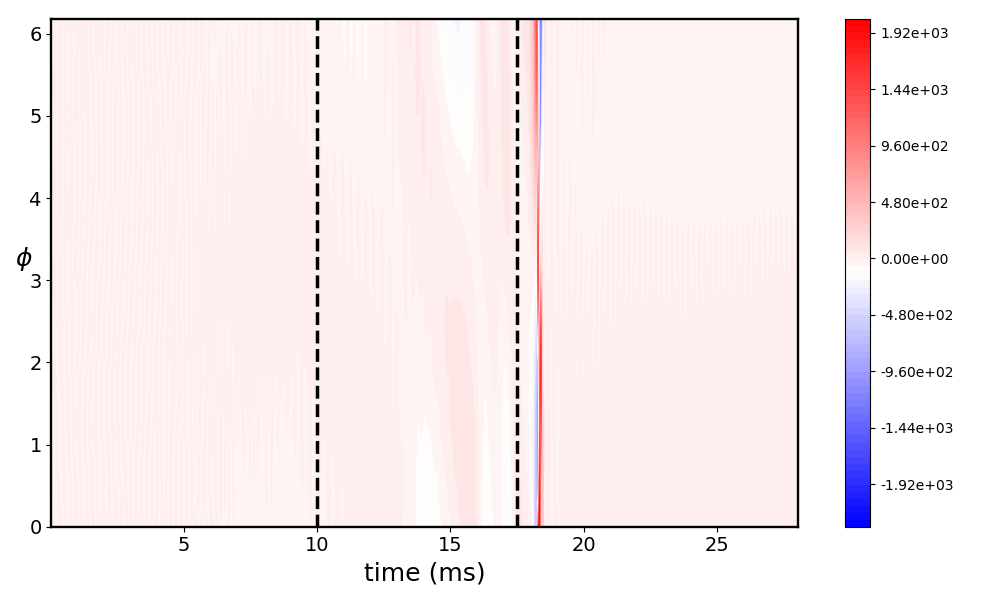}
		\caption{$\chi_{\parallel}/\chi_{\perp}= 10^{9}$}
		\label{fig_VDE_rect:asym_Ip_all_kpll1e9}
	\end{subfigure}
    \begin{subfigure}[htb]{0.49\textwidth}
		\centering
		\includegraphics[width=1\textwidth]{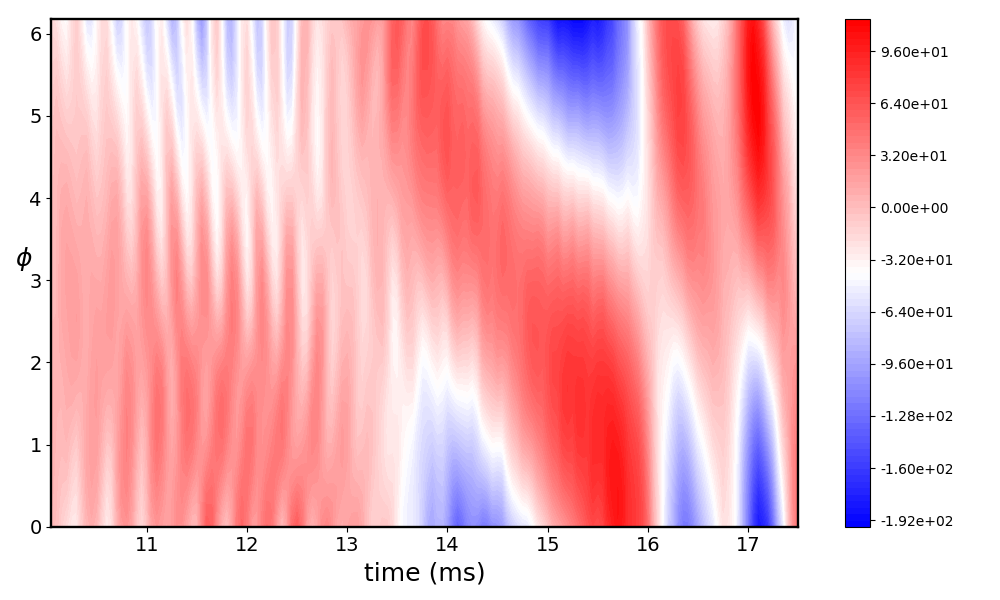}
		\caption{Zoomed-in time interval of (e)}
		\label{fig_VDE_rect:asym_Ip_zoom_kpll1e9}
	\end{subfigure}	

\caption{The time evolution of the toroidal asymmetry of plasma current $\widehat{I}_{p}(t,\phi)$ at each toroidal angle $\phi$ with (a) $\chi_{\parallel}/\chi_{\perp}= 10^{7}$, (c) $\chi_{\parallel}/\chi_{\perp}= 10^{8}$, and (e) $\chi_{\parallel}/\chi_{\perp}= 10^{9}$; the zoomed-in plots of (a), (c), and (e) for the corresponding selected time intervals are shown in (b), (d), and (f) respectively. }
\label{fig_VDE_rect:asym_Ip_scan_kpll}

\end{figure}
\begin{figure}[htb]
  \centering
  \includegraphics[width=1\textwidth]{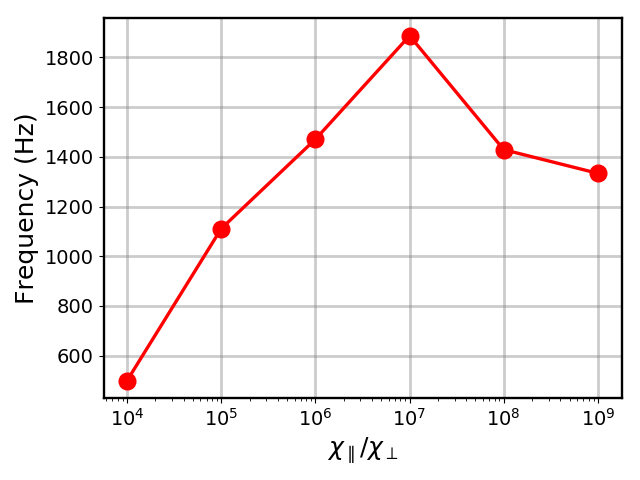}
  \caption{The dominant frequency in the temporal FFT spectrum of $\widehat{I}_{p}(t,\phi)$ along the toroidal direction from a non-axisymmetric VDE as a function of the ratio $\chi_{\parallel}/\chi_{\perp}$.}
  \label{fig_VDE_rect:scan_kpll_freq}
\end{figure}
\begin{figure}[htb]
  \centering
  \includegraphics[width=1\textwidth]{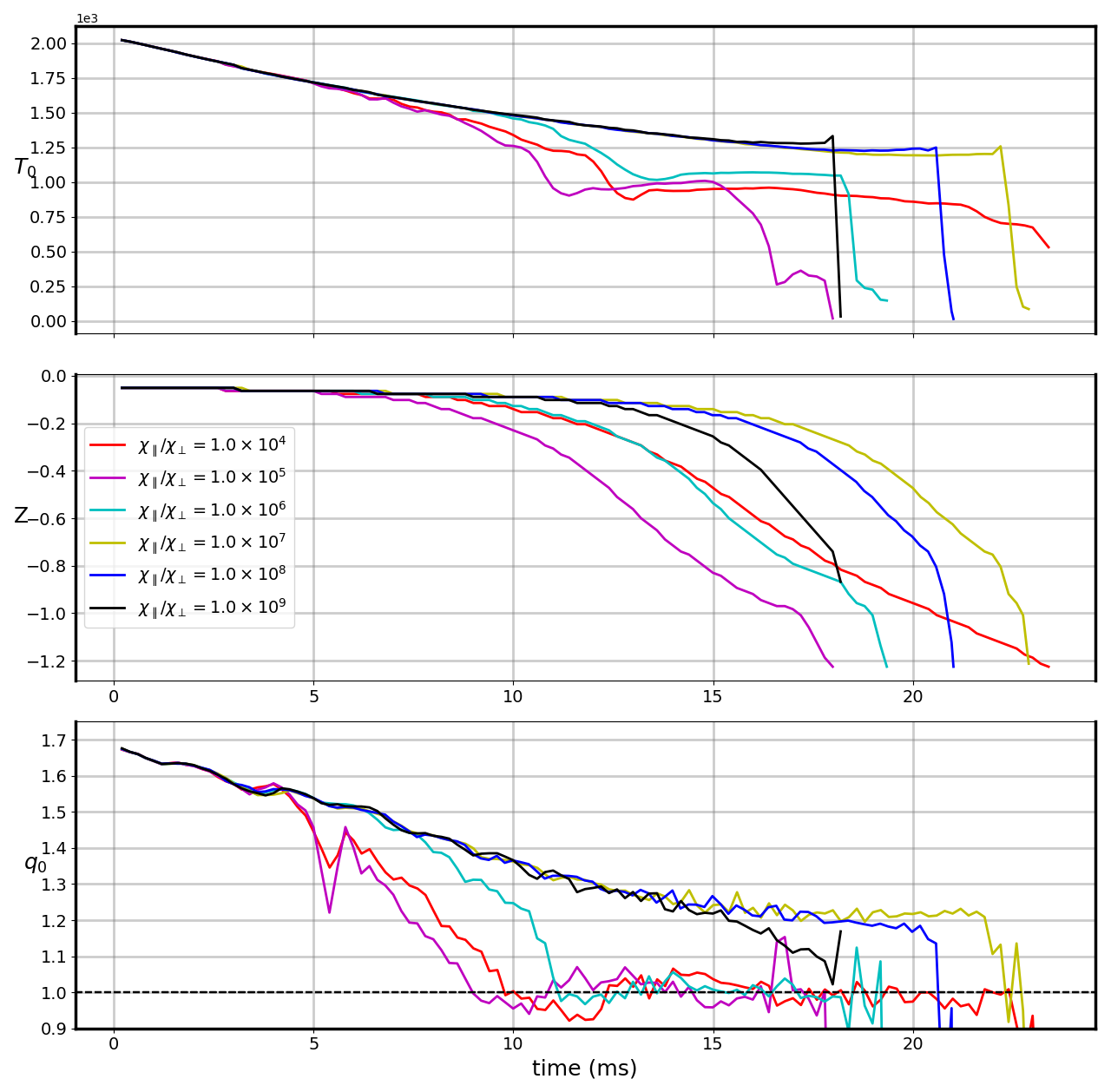}
  \caption{The time evolution of the plasma temperature $T_{0}$, the vertical coordinate $Z$, and the safety factor $q_{0}$ at magnetic axis during a non-axisymmetric VDE with various $\chi_{\parallel}/\chi_{\perp}$ values.}
  \label{fig_VDE_rect:3D_scan_kpll}
\end{figure}

The rotation frequency of the plasma current asymmetry increases with $\chi_{\parallel}/\chi_{\perp}$, where the toroidal mode $n=1$ always dominates (Fig.\ref{fig_VDE_rect:asym_Ip_scan_kpll}). The dominant frequency of the non-axisymmetric VDE changes with $\chi_{\parallel}/\chi_{\perp}$ and reaches maximum when $\chi_{\parallel}/\chi_{\perp}= 10^{7}$ (Fig.\ref{fig_VDE_rect:scan_kpll_freq}). There is no minor disruption and $q_{0}$ remains above unity when $\chi_{\parallel}/\chi_{\perp} \geq 10^{7}$. Since the characteristic width of magnetic island $W$ approaching thermal equilibration over the island region is proportional to $\left(\chi_{\parallel}/\chi_{\perp} \right)^{-1/4}$~\cite{1995PhPl}, the non-axisymmetric helical modes are suppressed with sufficiently large $\chi_{\parallel}/\chi_{\perp}$, so that the minor disruption disappears (Fig.\ref{fig_VDE_rect:3D_scan_kpll}). Note that the three dimensional physics during the VDE process are different when $\chi_{\parallel}/\chi_{\perp}$ lies below or above $10^{7}$. When $\chi_{\parallel}/\chi_{\perp} \textless 10^{7}$, the three dimensional perturbation mainly affects the VDE process through the formation of magnetic islands and the destruction of magnetic surfaces. When $\chi_{\parallel}/\chi_{\perp} \geq 10^{7}$, the minor disruption is suppressed and the larger $\chi_{\parallel}/\chi_{\perp}$ speeds up the TQ and hence the VDE process. Although the parallel thermal transport coefficient is large during the good confinement of tokamak plasma, it will soon decrease due to the rapid collapse of plasma temperature once the disruption or TQ process starts. Thus, the asymmetric effects in the lower regime of $\chi_{\parallel}/\chi_{\perp}$ may become more significant as the disruption progresses.

\section{Summary and Discussion}
\label{summary}

In summary, the non-axisymmetric VDE process has been studied in this work using NIMROD simulations. In comparison to the axisymmetric VDE simulation, during the non-axisymmetric VDE the magnetic island chains induced by the growth of the non-axisymmetric toridal modes stochastize the closed field lines from plasma boundary to the core region, which leads to jets of the poloidal current and plasma, and eventually the halo current as well as the finger and filamentary structures in the plasma pressure and temperature distributions.  The safety factor $q_{0}$ at the magnetic axis oscillates around unity, and the $(1,1)$ mode and island formation accelerates the collapse of core temperature and the advance of VDE process. The toroidal rotation frequency of the asymmetry in plasma current is close to that of the $q_{0}$ oscillation around unity and the non-axisymmetric perturbation has a dominant $n=1$ component, which suggest on the causal relationship between the $q_{0}$ oscillation and the rotating asymmetry in plasma current. 

Future work plans to focus on the roles of thermal transport in the destruction and healing of magnetic flux surfaces. The asymmetries in the wall force load distribution during a non-axisymmetric VDE will also be evaluated.


\section{Acknowledgment}

We are grateful for the helpful discussions with Prof. C.R. Sovinec, as well as the supports from the NIMROD team and the J-TEXT team. This work was supported by the National Key Research and Development Program of China (Grant Nos. 2019YFE03050004, 2022YFE03070000, and 2022YFE03070004), the National Natural Science Foundation of China (Grant No. 51821005), and the U.S. Department of Energy (Grant Nos. DE-FG02-86ER53218 and DE-SC0018001). We are grateful to Dr. Songtao Mao and Dr. Yanmin Duan of the EAST experimental group for providing the initial equilibrium data. The computing work in this paper is supported by the Public Service Platform of High Performance Computing by Network and Computing Center of HUST. This work was supported by Shenzhen Clean Energy Research Institute.

\newpage
\section{Reference}
\bibliographystyle{unsrt}
\bibliography{VDE}


\newpage
\newpage
\newpage
\newpage
\newpage

\newpage
\newpage
\newpage
\newpage

\newpage
\newpage

\newpage
\newpage
\newpage
\newpage
\newpage
\newpage
\newpage
\newpage
\newpage
\newpage

\newpage

\newpage
\newpage
\newpage
\newpage

\end{document}